\documentclass[acmsmall,screen]{acmart}

\AtBeginDocument{%
  }

\usepackage{array}
\usepackage{tabularx}
\usepackage{longtable}
\usepackage{colortbl}
\usepackage{pifont}
\usepackage{tcolorbox}

\definecolor{secgreen}{rgb}{0.00,0.60,0.30}
\definecolor{secred}{rgb}{0.80,0.00,0.00}
\definecolor{secyellow}{rgb}{0.71,0.51,0.00}
\definecolor{rowgray}{rgb}{0.96,0.96,0.96}
\definecolor{groupbg}{rgb}{0.86,0.86,0.86}
\newcommand{\cmark}{\textcolor{secgreen}{\ding{52}}}
\newcommand{\warn}{\textcolor{secyellow}{\ding{115}}}
\newcommand{\xmark}{\textcolor{secred}{\ding{56}}}
\newcommand{\crit}{\textcolor{secred}{\textbf{\ding{56}\ding{56}}}}
\newcommand{\rng}[2]{#1\,\text{--}\,#2}
\newcommand{\newmark}{\ding{108}}      
\newcommand{\partialmark}{\ding{119}}  
\newcommand{\nomark}{\ding{109}}       

\setcopyright{acmlicensed}
\copyrightyear{2026}
\acmYear{2026}
\acmDOI{XXXXXXX.XXXXXXX}

\acmJournal{JACM}
\acmVolume{37}
\acmNumber{4}
\acmArticle{111}
\acmMonth{8}

\begin{document}

\title{Towards Secure Agent Skills: Architecture, Threat Taxonomy, and Security Analysis}

\author{Zhiyuan Li}
\email{lizhiyuan2021@iscas.ac.cn}
\orcid{0009-0000-0001-7097}
\authornote{Institute of Software, Chinese Academy of Sciences, Beijing, China}
\authornote{University of Chinese Academy of Sciences, Beijing, China}

\author{Jingzheng Wu}
\email{jingzheng08@iscas.ac.cn}
\orcid{0000-0001-5561-9829}
\authornotemark[1] 
\authornote{Key Laboratory of System Software (Chinese Academy of Sciences), Beijing, China}
\authornote{Jingzheng Wu is the corresponding author.}

\author{Xiang Ling}
\email{lingxiang@iscas.ac.cn}
\orcid{0000-0002-7377-7844}
\authornotemark[1]
\authornotemark[3]  

\author{Xing Cui}
\email{cuixing@iscas.ac.cn}
\orcid{0000-0002-0810-562X}
\authornotemark[1]
\authornotemark[2]

\author{Tianyue Luo}
\email{tianyue@iscas.ac.cn}
\orcid{0000-0001-7407-8255}
\authornotemark[1]

\renewcommand{\shortauthors}{Li et al.}

\begin{abstract}
Agent Skills is an emerging open standard that defines a modular, filesystem-based packaging format enabling LLM-based agents to acquire domain-specific expertise on demand. Despite rapid adoption across multiple agentic platforms and the emergence of large community marketplaces, the security properties of Agent Skills have not been systematically studied. This paper presents a systematization of the security of the Agent Skills framework. We define a threat model for the framework and decompose the full lifecycle of an Agent Skill across four phases---Creation, Distribution, Deployment, and Execution---identifying the structural attack surface each phase introduces. Building on this lifecycle analysis, we construct a threat taxonomy comprising seven categories and nineteen scenarios organized across three attack layers, grounded in both architectural analysis and real-world evidence. Each scenario is annotated with whether the exposure follows from the specification or from the latitude the specification leaves to individual runtimes, with the strength of the public evidence supporting it, and with the feasibility of its mitigation. We map seven documented security incidents onto the taxonomy. Based on these findings, we discuss defense directions for each threat category, identify open research challenges, and provide actionable recommendations for stakeholders. The most severe threats arise from structural properties of the framework itself: the absence of a data-instruction boundary, a single-approval persistent trust model, and a capability declaration mechanism to which the specification attaches no enforcement semantics. None of these is addressable through incremental mitigations alone.
\end{abstract}



\maketitle

\section{Introduction}

The rapid advancement of AI agents based on large language models (LLMs) has fundamentally transformed how humans interact with software systems~\cite{xi2025rise,wang2024survey}. Modern LLM-based agents are no longer confined to passive question-answering. They actively plan multi-step workflows, execute code, and interact with external services with minimal human oversight. This shift toward agentic AI has driven the emergence of capability extension frameworks that allow agents to acquire domain-specific expertise on demand, enabling specialization across an unbounded range of tasks without retraining the underlying model.

Among these frameworks, \textit{Agent Skills}---introduced by Anthropic in October 2025---represents a significant architectural departure from prior approaches such as ChatGPT Plugins~\cite{openai2023plugins} and the Model Context Protocol (MCP)~\cite{hou2025mcp}. Rather than exposing typed API schemas or defining a typed invocation protocol, the Agent Skills framework organizes capabilities as modular, filesystem-based directories. Each Skill bundles a \texttt{SKILL.md} instruction file written in natural language, optional executable scripts, and reference resources, which the agent loads on demand when it determines the Skill to be relevant to the user's request~\cite{anthropic2025skills}. This design achieves remarkable flexibility and composability, as Skills can encode arbitrary workflows, domain knowledge, and organizational procedures in a format that requires no programming expertise to author. Adoption followed quickly. Anthropic released the specification as an open standard in December 2025 and transferred its stewardship to the Agentic AI Foundation under the Linux Foundation~\cite{agentskills2025spec,aaif2025}, and the format has since been adopted by several dozen agent products, including Cursor~\cite{cursor2025skills}, GitHub Copilot~\cite{github2025skills}, Gemini CLI~\cite{google2026geminicli}, and OpenAI Codex~\cite{agentskills2026showcase}. Community marketplaces aggregating Skills at a scale of hundreds of thousands of packages emerged over the same period, and the major ones impose no mandatory security review.

This rapid adoption has outpaced the development of adequate security mechanisms. The architectural properties that make Skills powerful---natural-language instruction delivery, filesystem-level code execution, and open marketplace distribution---create security vulnerabilities that are qualitatively distinct from those of prior AI extension mechanisms. Real-world incidents confirm that these risks are not theoretical. In December 2025, security researchers demonstrated the execution of live ransomware via a weaponized Agent Skill. The attack exploited a \textit{consent gap}: once a user approves a Skill, it silently inherits persistent permissions to read and write files, download code, and open network connections, all without further prompts~\cite{cherny2025medusalocker}. In February 2026, a coordinated supply chain campaign compromised at least 1,184 Skills in a major community marketplace, delivering credential-theft payloads to unsuspecting users~\cite{liu2026malicious,antiy2026clawhavoc}. A large-scale empirical study published in January 2026, which scanned 42,447 Skills, found that 26.1\% contained at least one security vulnerability, spanning 14 distinct patterns across four categories: prompt injection, data exfiltration, privilege escalation, and supply chain risks~\cite{liu2026agent}. Nor has the pressure eased as the ecosystem matured: a campaign that ran throughout July 2026 accumulated more than 1.7 million installations before it was disrupted, harvesting SSH keys, cloud credentials, and package-manager tokens from developer workstations and CI runners~\cite{zenity2026skillsupplychain}. Independent researchers further demonstrated that Skill-based prompt injection constitutes a qualitatively harder attack class than conventional indirect injection, because Skill files are composed entirely of instructions with no data-to-instruction boundary~\cite{schmotz2025trivial,schmotz2026skillinject}.

These incidents are not isolated failures attributable to implementation oversights. They reflect \emph{structural} properties of the Agent Skills framework: a trust model that treats instructions as operator-level directives; a consent mechanism that grants persistent permissions from a single approval; a distribution model that imposes no mandatory security review; and a runtime model that executes bundled scripts with the user's local privileges. Despite the severity and breadth of these issues, the security implications of these properties have so far been examined only in fragments. Vendor reports document individual campaigns, measurement studies quantify the prevalence of particular vulnerability patterns, and attack papers demonstrate single techniques in isolation, but no account connects them---one that traces which architectural decision produces which attack surface, at which point in a Skill's lifecycle that surface becomes exploitable, and which of the resulting threats admit incremental mitigation as opposed to requiring reform of the specification itself. Practitioners deploying Skills consequently have little principled guidance beyond the vendor's recommendation to \textit{``only install Skills from trusted sources''}~\cite{useskills}.

In this paper, we present a systematization of the security of the Agent Skills framework. Our goal is to characterize its structural security properties---the threat model it creates, the attack surfaces it exposes across its full lifecycle, and the research challenges that must be addressed to make Skills safe for broad deployment. Our analysis rests on public evidence: the published specification and platform documentation, disclosed incidents and CVE records, and peer-reviewed and preprint measurement studies. We make the following contributions:

\begin{itemize}
  \item \textbf{Threat model and lifecycle analysis.} We define a threat model for the Agent Skills framework---the attacker's capabilities and objectives, the trust assumptions the framework makes, and the boundaries of our analysis---and decompose the framework into four security-relevant phases: \emph{Creation}, \emph{Distribution}, \emph{Deployment}, and \emph{Execution}. We identify the structural security implications of each phase (Section~\ref{sec:lifecycle}).
  \item \textbf{Threat taxonomy.} We construct a threat taxonomy for the Agent Skills framework, identifying 7 threat categories and 19 distinct threat scenarios across three attack layers. Each scenario is annotated with the strength of the evidence supporting it, with the feasibility class of its mitigation, and with whether it originates in the specification itself or in the latitude the specification leaves to individual runtimes (Section~\ref{sec:taxonomy}).
  \item \textbf{Incident analysis.} We analyze seven documented security incidents in the Agent Skills ecosystem under explicit selection criteria and map each to the taxonomy (Section~\ref{sec:incidents}).
  \item \textbf{Defense directions and research agenda.} We discuss mitigation strategies for each threat category and identify eight open research challenges that must be addressed to establish Agent Skills security as a mature research area (Section~\ref{sec:discussion}).
\end{itemize}

Our analysis reveals that the Agent Skills attack surface spans seven threat categories organized across three attack layers. The first layer covers how malicious Skills reach users and acquire trusted authority, encompassing \emph{supply chain compromise}, facilitated by open marketplaces with no mandatory vetting, and \emph{consent abuse}, arising from the single-approval persistent trust model. The second layer covers direct attacks that an activated Skill can mount, including \emph{prompt injection}, enabled by the absence of a structural data-to-instruction boundary; \emph{code execution}, exploiting bundled scripts and runtime dependency mechanisms; and \emph{data exfiltration} through credential harvesting, environment variable access, and silent codebase transmission. The third layer covers how compromise effects extend beyond the current session or agent boundary, through \emph{persistence} via memory file and configuration poisoning, and \emph{multi-agent propagation} in orchestrated agentic pipelines. Across all three layers, the threats that resist mitigation are concentrated in those that follow from the specification rather than from any individual implementation: addressing them requires not only improved tooling and marketplace governance, but architectural reform of the Agent Skills framework itself.

The remainder of this paper is organized as follows. Section~\ref{sec:background} provides background on prior agent capability extension mechanisms. Section~\ref{sec:architecture} presents the architecture of the Agent Skills framework. Section~\ref{sec:lifecycle} defines our threat model and analyzes the security implications of each lifecycle phase. Section~\ref{sec:taxonomy} develops our threat taxonomy. Section~\ref{sec:incidents} analyzes real-world incidents. Section~\ref{sec:discussion} discusses defense directions and open challenges. Section~\ref{sec:related} surveys related work and positions our taxonomy against concurrent studies. Section~\ref{sec:conclusion} concludes.

\section{Background}\label{sec:background}

Agent Skills is the third major attempt to give language models capabilities beyond their training corpus. Each of the three mechanisms resolved a limitation of its predecessor, and each resolution traded constraint for expressiveness. Section~\ref{sec:background:plugins} to Section~\ref{sec:background:skills} describe the three mechanisms in turn, and Section~\ref{sec:background:comparison} compares them along the dimensions that determine their security posture.

\subsection{ChatGPT Plugins}\label{sec:background:plugins}

ChatGPT Plugins (2023--2024) were introduced to overcome a fundamental limitation of language models: their inability to access real-time information and third-party services beyond the training corpus~\cite{openai2023plugins}. Each plugin exposed a typed API manifest in OpenAPI format, which the model used to construct well-formed HTTP requests to a remote, operator-controlled endpoint. This architecture extended the model's reach to live data and external services while preserving a strong security boundary. Execution occurred entirely on the plugin provider's server infrastructure, with no local code running on the user's machine. The schema contract further constrained the action space, allowing only operations defined in the manifest with typed parameters. Users enabled plugins on a per-conversation basis, and the authority a plugin held was bounded by its manifest. Plugins were subject to a mandatory review process combining automated and human checks before publication in the plugin store~\cite{customgpt2023review}.

The mechanism was nonetheless short-lived. The need for programming expertise to author an OpenAPI specification, the restriction to typed remote API calls, and the inability to encode complex multi-step procedural workflows led to low developer and user adoption~\cite{datacamp2024plugins}. OpenAI stopped accepting new plugin conversations on March 19, 2024 and shut the mechanism down entirely on April 9, 2024, redirecting developers to custom GPTs and later to MCP-based connectors~\cite{openai2024sunset}. Plugins remains the only widely deployed extension mechanism whose manifest bound the capability's action space to a declared and reviewable set, and it serves as the reference point against which the security regressions of its successors are measured.

\subsection{Model Context Protocol}\label{sec:background:mcp}

The Model Context Protocol (MCP, November 2024) was designed to solve a different problem: the M$\times$N integration explosion that arose as AI systems began connecting to diverse external tools and data sources~\cite{anthropic2024mcp}. Before MCP, connecting $M$ AI applications to $N$ tools required up to $M \times N$ custom integrations. MCP replaced this with a universal JSON-RPC 2.0 protocol that any compliant client could use to discover and invoke capabilities exposed by any compliant server~\cite{hou2025mcp}. This standardization dramatically reduced integration overhead and enabled a thriving ecosystem of reusable server implementations. Governance of the protocol was subsequently transferred to the Agentic AI Foundation~\cite{aaif2025}. Servers declare tools, resources, and prompts with structured schemas. This typed interface also preserved a partial data-to-instruction boundary, because the model invokes typed operations with well-defined parameters rather than interpreting free-form natural language directives.

MCP nonetheless introduced new security trade-offs relative to Plugins. MCP servers may run locally with user-level system access, broadening the attack surface. More significantly, MCP adopted a fully decentralized distribution model with no mandatory review process. Servers are typically installed directly from source repositories, eliminating the centralized vetting checkpoint that the plugin store provided~\cite{hou2025mcp}.

\subsection{Agent Skills}\label{sec:background:skills}

Agent Skills (October 2025) addressed limitations that remained even after MCP: the difficulty of encoding complex procedural workflows, organizational context, and domain-specific knowledge in a typed schema interface, and the barrier to authorship that requiring an MCP server implementation imposed~\cite{anthropic2025skills}. By replacing the typed interface with a natural language instruction file (\texttt{SKILL.md}), the Agent Skills framework made it possible for anyone to package expertise into a reusable agent capability. The progressive disclosure design further reduced the token overhead of maintaining large capability libraries.

These design choices introduced security trade-offs that are qualitatively more severe than those of either protocol. The co-location of natural language instructions and executable scripts within a single filesystem package collapses the distinction between capability specification and code execution. A \texttt{SKILL.md} file may direct the agent to run a bundled script that in turn downloads and executes arbitrary network-hosted code. The single-approval persistent trust model positions an activated Skill's instructions at operator level from a single installation event, without the per-action confirmation that MCP clients may enforce. The major community marketplaces, like MCP repositories but unlike the plugin store, impose no mandatory security review before publication~\cite{snyk2026toxicskills}.

\subsection{Comparative Analysis}\label{sec:background:comparison}

\begin{table}[t]
\centering
\small
\setlength{\tabcolsep}{4pt}
\renewcommand{\arraystretch}{1.3}
\caption{Comparison of agent capability extension mechanisms across security-relevant dimensions.}
\label{tab:comparison}
\begin{tabular}{
  >{\raggedright\arraybackslash}m{\dimexpr 0.24\linewidth-2.4\tabcolsep\relax}
  >{\raggedright\arraybackslash}m{\dimexpr 0.36\linewidth-3.6\tabcolsep\relax}
  >{\centering\arraybackslash}m{\dimexpr 0.15\linewidth-1.5\tabcolsep\relax}
  >{\centering\arraybackslash}m{\dimexpr 0.1\linewidth-1.0\tabcolsep\relax}
  >{\centering\arraybackslash}m{\dimexpr 0.15\linewidth-1.5\tabcolsep\relax}
}
\toprule
\textbf{Dimension}
  & \textbf{Description}
  & \textbf{\shortstack{ChatGPT\\Plugins}}
  & \textbf{MCP}
  & \textbf{\shortstack{Agent\\Skills}} \\
\midrule

\rowcolor{groupbg}
\multicolumn{5}{l}{\textit{\small Specification-Determined Properties}} \\

Instruction Carrier
  & Format used to convey the capability specification to the agent
  & \cmark & \warn & \xmark \\
\rowcolor{rowgray}
Data/Instruction Boundary
  & Whether instructions and runtime data are structurally separated
  & \cmark & \warn & \crit \\
Capability Declaration
  & Whether a capability must declare the resources it accesses, and whether that declaration binds
  & \cmark & \warn & \xmark \\
\rowcolor{rowgray}
Trust Model
  & Granularity and persistence of the authority granted when a capability is enabled
  & \cmark & \xmark & \crit \\

\midrule

\rowcolor{groupbg}
\multicolumn{5}{l}{\textit{\small Implementation-Determined Properties}} \\

Execution Locus
  & Where capability code executes relative to the user's machine
  & \cmark & \rng{\warn}{\xmark} & \rng{\warn}{\xmark} \\
\rowcolor{rowgray}
Runtime Isolation
  & Degree of sandboxing applied to capability execution
  & \cmark & \rng{\warn}{\xmark} & \rng{\warn}{\crit} \\
Permission Scope
  & Breadth of system resources accessible during execution
  & \cmark & \rng{\warn}{\xmark} & \rng{\warn}{\crit} \\

\midrule

\rowcolor{groupbg}
\multicolumn{5}{l}{\textit{\small Distribution \& Ecosystem Governance}} \\

Marketplace Review
  & Whether mandatory vetting exists before public distribution
  & \cmark & \xmark & \xmark \\
\rowcolor{rowgray}
Authorship Complexity
  & Technical expertise required to create and publish a capability
  & High & Medium & Low \\

\bottomrule
\end{tabular}

\vspace{3pt}
\begin{minipage}{\linewidth}\footnotesize
\cmark~mitigated, \warn~partially exposed, \xmark~exposed, \crit~critically exposed.
Assignment criteria are defined in Section~\ref{sec:background:comparison}, and the principles governing the assignments together with the basis for each cell are recorded in~\ref{app:rubric}.
Implementation-determined dimensions are left unspecified by the respective standard, and ranges give the span across observed runtimes.
Authorship Complexity uses a separate scale, the barrier to creating and publishing a capability, and is not a security level. A lower barrier increases supply chain risk.
\end{minipage}
\end{table}

Table~\ref{tab:comparison} compares the three mechanisms along nine dimensions. We derive these dimensions from the lifecycle decomposition developed in Section~\ref{sec:lifecycle}, taking for each phase the design decisions that determine what an adversary can do at that phase. The Creation phase contributes the instruction carrier and the capability declaration mechanism, since these fix what an author can express and what they must disclose. The Distribution phase contributes marketplace review and authorship complexity, which jointly determine the cost of placing a malicious capability in front of users. The Deployment phase contributes the trust model, which fixes what a user grants when a capability is enabled. The Execution phase contributes the data/instruction boundary, execution locus, runtime isolation, and permission scope, which bound what an activated capability can reach.

We assign each cell one of four levels. A dimension is marked \cmark{} where the design prevents the exposure by construction, so that an adversary cannot exercise it without violating the mechanism's own contract, and \warn{} where the exposure is constrained but enforcement depends on correct client behavior. Where no enforcement exists and risk is bounded only by convention or user vigilance, we mark \xmark{}. The mark \crit{} additionally requires that the design amplify the exposure, by making it the default, by concealing it from the user, or by granting it elevated authority. An amplifying property is counted once, at the dimension that produces it, rather than at every dimension it affects. The full set of principles and the basis for each cell are recorded in~\ref{app:rubric}.

The table groups the dimensions by what fixes them. Specification-determined properties hold across all conforming implementations; implementation-determined properties are left open by the standard, so each runtime resolves them independently, and the ranges span the implementations we observed. The distinction has direct consequences for how security claims about Agent Skills should be scoped. The same \texttt{SKILL.md} executes under sandbox isolation on the Claude API and with full local host privileges under Claude Code~\cite{pluto2026extensions}, and Claude Code itself now offers filesystem and network isolation modes that were absent at the format's introduction~\cite{anthropic2026sandbox}. Claims about execution locus, isolation, and permission scope are therefore claims about a particular deployment rather than about Agent Skills as such.

The security regression from Plugins to Agent Skills is concentrated where it is least remediable. Plugins is mitigated on every specification-determined dimension, whereas Agent Skills is exposed or critically exposed on all four. The implementation-determined dimensions admit better and worse resolutions, and the range in the Agent Skills column reflects runtimes that have already chosen differently. What no individual cell conveys is how the two groups interact. A runtime may confine a Skill to a sandbox, yet neither the user nor an auditing tool can tell whether that confinement is adequate, because the Skill was never required to state what it needs. The absence of a binding capability declaration is what leaves every isolation decision unverifiable. This coupling, rather than the severity of any single dimension, is what distinguishes Agent Skills from its predecessors. Section~\ref{sec:architecture} examines the Agent Skills architecture that produces these properties in detail.

\section{The Architecture of Agent Skills}\label{sec:architecture}

Agent Skills define a filesystem-based packaging format that enables LLM-based agents to acquire domain-specific expertise on demand, without retraining the underlying model~\cite{anthropic2025skills}. The format was released as an open standard in December 2025 and is now governed by the Agentic AI Foundation~\cite{agentskills2025spec,aaif2025}. Its security properties follow from three interlocking components: the package structure that defines what a Skill contains, the progressive disclosure model that governs when and how Skill content is activated, and the trust model that determines what authority an activated Skill holds.

\subsection{Package Structure}\label{sec:architecture:package}

Each Agent Skill is a self-contained directory. The only required file is \texttt{SKILL.md}, which serves as the agent-facing entry point~\cite{anthropic2025skills}. The Skill directory may additionally include executable scripts in any supported runtime such as Python, Bash, or JavaScript, reference assets such as data files, templates, and configuration files, and supplementary instruction files that \texttt{SKILL.md} references by name.

\begin{table}[t]
\centering
\caption{Representative structure of a \texttt{SKILL.md} file.}
\label{tab:skillmd}
\begin{tcolorbox}[colback=gray!10, colframe=gray!50, arc=2mm, boxrule=0.5pt, left=3mm, right=3mm, top=1mm, bottom=1mm]
\begin{tabularx}{\linewidth}{X}
\textbf{Frontmatter (YAML header, required):}\\
\texttt{-{}-{}-}\\
\texttt{name: pdf-processing}\\
\texttt{description: Extract text and tables from PDF files. Use when working}\\
\texttt{\phantom{xxxxxxxxxxxxx}with PDF files or when the user mentions PDFs.}\\
\texttt{-{}-{}-}\\
\textbf{Instructions body (Markdown, natural language):}\\
\texttt{\#\# Instructions}\\
When the user asks to process a PDF file, use the bundled \texttt{extract.py} script to extract its contents:\\
\texttt{\phantom{xx}python extract.py -{}-input <file> -{}-output <dir>}\\
You may read any PDF from the current working directory and write results to the output directory. If the user requests merging, call \texttt{merge.py} with the list of input files. For network-hosted PDFs, download them first using the fetch tool before processing.\\
\end{tabularx}
\end{tcolorbox}
\end{table}

The \texttt{SKILL.md} file follows a two-part structure~\cite{anthropic2025skills}. The \emph{frontmatter} is a YAML header block delimited by ``\texttt{-{}-{}-}" markers. It declares two required metadata fields: \texttt{name}, a short identifier for the Skill, and \texttt{description}, a natural language summary that the agent uses to assess whether the Skill is relevant to the current task. The \emph{instructions body} is an unstructured natural language section written in Markdown that provides the agent with behavioral directives for the Skill's operation. This section may reference bundled supplementary files, direct the agent to execute bundled scripts, invoke external network resources, or read and write local files. Table~\ref{tab:skillmd} shows a representative \texttt{SKILL.md} structure.

The frontmatter also admits optional fields, several of which bear on security. The \texttt{allowed-tools} field names the tools a Skill may invoke, and \texttt{context: fork} directs a runtime to execute the Skill in a separate context window rather than the main conversation~\cite{agentskills2025spec}. The specification attaches no enforcement semantics to either. It states what an author may declare, not what a runtime must do with the declaration. Conforming runtimes have consequently diverged: Claude Code applies \texttt{allowed-tools} when authorizing tool use, whereas the Anthropic Agent SDK does not read the field at all, so the same Skill runs with the tool set the host application grants rather than the one its author specified~\cite{anthropic2026skillsbestpractices}. A declaration that binds under one runtime and is silently discarded under another cannot be relied upon by a user deciding whether to install a Skill, nor by a tool attempting to audit one.

Two properties of \texttt{SKILL.md} follow from this design. The first is the absence of a formal interface contract between the frontmatter and the instructions body. The \texttt{description} field declares the Skill's stated purpose, but no mechanism verifies that the instructions body is consistent with that declaration or confined to it. The second is the absence of any separation between instructions authored by the Skill developer and data that the Skill consumes at runtime. Both are natural language text within the same document, and the language model processes them identically. Mechanisms that defend against injection in other settings depend on exactly this separation. Parameterized SQL queries prevent injection because the query template and the parameter values travel through distinct channels, so a value can never be reinterpreted as part of the query. Agent Skills provide no equivalent channel distinction, and consequently no basis on which a runtime could refuse to execute retrieved text as a directive. Section~\ref{sec:taxonomy} analyzes the attacks this permits.

\subsection{Progressive Disclosure Loading Model}\label{sec:architecture:loading}

\begin{figure}[t]
  \centering
  \includegraphics[width=\columnwidth]{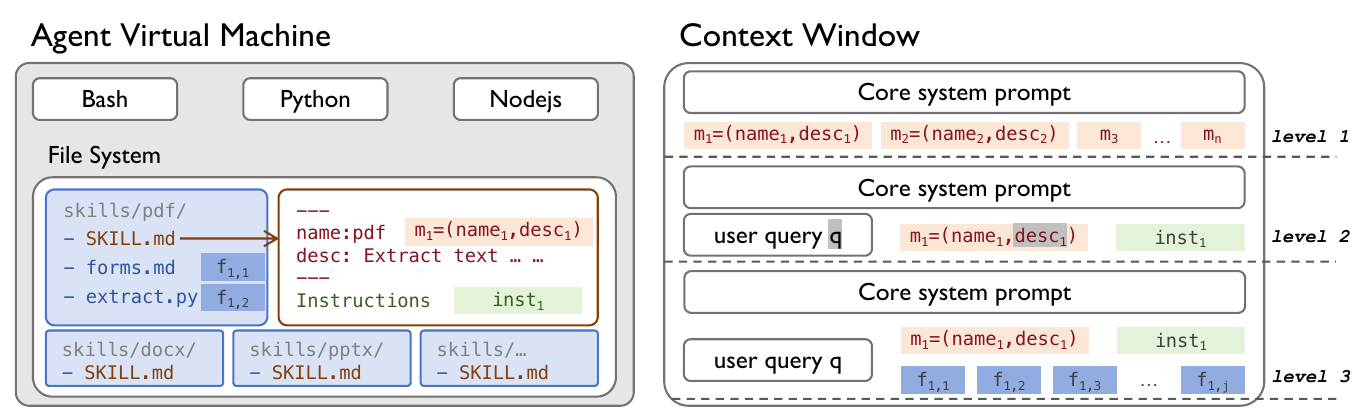}
  \caption{The Agent Skills architecture. Left: the filesystem layout of a Skill package within the agent's virtual machine, comprising \texttt{SKILL.md}, supplementary instruction files, and executable scripts. Right: the three-level progressive disclosure process, showing how Skill metadata ($m_i$), instructions ($\mathit{inst}_i$), and supplementary files ($f_{i,j}$) are loaded into the context window incrementally across \emph{Levels 1, 2, and 3}.}
  \label{fig:3level}
  \Description{The Agent Skills architecture. Left: the filesystem layout of a Skill package within the agent's virtual machine, comprising \texttt{SKILL.md}, supplementary instruction files, and executable scripts. Right: the three-level progressive disclosure process, showing how Skill metadata ($m_i$), instructions ($\mathit{inst}_i$), and supplementary files ($f_{i,j}$) are loaded into the context window incrementally across \emph{Levels 1, 2, and 3}.}
\end{figure}

The Agent Skills runtime activates Skill content through a three-level progressive disclosure process, designed to scale across large Skill repositories without exhausting the agent's context window~\cite{anthropic2025skills}. Let $\mathcal{S} = \{s_1, s_2, \ldots, s_n\}$ denote the set of Skills installed in the agent's environment, and let each Skill $s_i$ be characterized by a metadata pair $m_i = (\mathit{name}_i, \mathit{desc}_i)$ extracted from its frontmatter, a full instruction document $\mathit{inst}_i$ comprising the \texttt{SKILL.md} instructions body, and an optional set of supplementary files $\mathit{files}_i = \{f_{i,1}, f_{i,2}, \ldots\}$ bundled within the Skill directory. Figure~\ref{fig:3level} illustrates the overall architecture, showing the filesystem layout of a Skill package on the left and the three-level progressive disclosure process within the agent's context window on the right.

At \emph{Level 1}, the agent pre-loads the metadata pair $m_i$ for all $s_i \in \mathcal{S}$ at startup, injecting each Skill's \texttt{name} and \texttt{description} directly into the agent's system prompt. This provides just enough information for the agent to reason about Skill relevance without loading any instruction content into the context window.

At \emph{Level 2}, given a user request $q$, the agent evaluates a relevance function $r(q, m_i) \in \{0,1\}$ for each $s_i$, where $r(q, m_i) = 1$ if the agent determines that $s_i$ is applicable to $q$. This determination is made by the language model itself through semantic reasoning over $\mathit{desc}_i$. For each $s_i$ with $r(q, m_i) = 1$, the agent retrieves $\mathit{inst}_i$ by invoking a Bash tool to read \texttt{SKILL.md} from the filesystem, bringing its full content into the active context window. Because $r$ operates over author-supplied natural language rather than over a registry of verified identifiers, an author controls both operands of the selection decision. Section~\ref{sec:taxonomy} examines the consequences for Skill selection.

At \emph{Level 3}, if $\mathit{inst}_i$ references specific supplementary files $f_{i,j} \in \mathit{files}_i$, the agent reads those files on demand using additional Bash commands, loading only the subset of supplementary content relevant to the current task. When $\mathit{inst}_i$ directs the agent to execute a bundled script, the agent runs the script via Bash. The script's source code never enters the context window, and only the script's output is returned to the agent~\cite{anthropic2025skills}. The agent sees a result without seeing what produced it.

Let $\mathcal{A}$ denote the set of actions available to the agent, comprising filesystem operations, network calls, and subprocess invocations. Let $\mathcal{D}_i \subseteq \mathcal{A}$ denote the action set that $s_i$ declares, taken as the union of what $\mathit{desc}_i$ states and what the optional frontmatter fields name. Because the instructions body is interpreted directly by the language model with no enforcement mechanism, a Skill can in principle direct the agent to perform any action in $\mathcal{A}$, not merely those in $\mathcal{D}_i$. The specification neither requires $\mathcal{D}_i$ to be complete nor obliges a runtime to confine execution to it, which leaves $\mathcal{A} \setminus \mathcal{D}_i$ available to every installed Skill and makes it the quantity the trust model governs.

\subsection{Trust Model and Permission Scope}\label{sec:architecture:trustmodel}

The specification positions an installed and activated Skill at operator level. Operator-level instructions are incorporated into the agent's system prompt and are documented as carrying elevated authority that the agent is not designed to override on the basis of subsequent user input~\cite{anthropic2025skills}. By injecting $m_i$ into the system prompt at startup and subsequently treating $\mathit{inst}_i$ as an extension of that system-prompt context, the framework positions each activated Skill to issue instructions that the agent will follow without per-action user confirmation.

Skill instructions enter the context with the standing of system-prompt content, and nothing marks them as untrusted. How a given model resolves a conflict between such an instruction and a contradicting user request depends on that model and on the runtime hosting it, and reported behavior varies across models under identical Skill-borne attacks~\cite{schmotz2026skillinject}.

The \emph{consent gap} is the mismatch between the permission object at approval time and the permission scope at execution time~\cite{cherny2025medusalocker}. When a user installs a Skill, the object of consent is the Skill's declared \texttt{description} and, if reviewed, the content of \texttt{SKILL.md} at that moment. The scope actually granted is $\mathcal{A}$ rather than $\mathcal{D}_i$, and it holds for all future sessions without per-action or per-session confirmation. Three properties compound this gap. First, the grant is \emph{persistent}. No re-approval is required when the Skill is invoked in a new session. Second, the grant is \emph{undifferentiated}. A single installation approval covers all actions the Skill may subsequently direct, regardless of their scope or reversibility. Third, the grant is \emph{irrevocable by content change}. An author or a supply chain attacker who modifies $\mathit{inst}_i$ after installation inherits the original approval, because the trust relationship is bound to the Skill identity rather than to a committed version of its content.

The gap extends beyond the agent's own action space, because agents in current deployments read behavioral guidance from persistent files in the working directory and the user profile. Files such as \texttt{AGENTS.md} and \texttt{CLAUDE.md} are loaded at startup and carry standing instructions across sessions~\cite{agentsmd2025}. These files fall within $\mathcal{A}$ for any Skill that can write to the filesystem, and a Skill that modifies them alters agent behavior in sessions where the Skill itself is never activated. Section~\ref{sec:taxonomy} treats this as a distinct threat category.

Two platform mechanisms bear on the gap. The optional frontmatter fields of Section~\ref{sec:architecture:package} let an author state a narrower $\mathcal{D}_i$, and Claude Code offers filesystem and network isolation modes that bound $\mathcal{A}$ itself~\cite{anthropic2026sandbox}. Both are opt-in rather than required by the specification, and neither binds the trust grant to a version of a Skill's content. Official user-facing guidance continues to rest on the recommendation to ``only install Skills from trusted sources''~\cite{useskills}.

\section{Threat Model and Lifecycle Attack Surface}\label{sec:lifecycle}

\subsection{Threat Model}\label{sec:lifecycle:threatmodel}

We consider an adversary whose objective is to cause an agent to act against the interests of the user in whose environment it runs. Concretely, the adversary seeks to execute code on the user's host, to read and transmit data the user has not authorized for disclosure, to alter agent behavior in sessions beyond the one under attack, or to redirect the agent's actions toward attacker-chosen ends while the user believes a legitimate task is being performed.

\textbf{Adversary capabilities.} The adversary may author Skills and publish them to any marketplace that accepts submissions, and may do so under an identity of their choosing, since no marketplace binds publication to a verified identity. The adversary may modify a Skill's content after publication and after users have installed it. The adversary may register package names on public registries such as PyPI and npm, and may control the content served at any URL a Skill fetches at runtime. The adversary may also obtain control of a legitimate Skill's distribution channel through credential theft or account takeover, in which case they inherit that Skill's existing installation base. Where a marketplace ranks Skills by download counts or ratings, the adversary may inflate those signals.

\textbf{Adversary limitations.} The adversary cannot modify the underlying language model or its training data, cannot break cryptographic primitives, and has no privileged access to the user's machine beyond what an installed Skill obtains. The adversary cannot compel a user to install a specific Skill, and must instead rely on the user's or the agent's selection among available Skills.

\textbf{Trust assumptions.} The framework as specified places trust in three places. It trusts that a Skill's declared \texttt{description} describes its behavior, because the description is the sole basis on which both the user's installation decision and the agent's relevance determination are made. It trusts the contents of the Skill directory on the filesystem, since the runtime reads \texttt{SKILL.md}, supplementary files, and bundled scripts without verifying their provenance against the version the user approved. And it trusts a single installation approval as authorization for every subsequent activation.

\textbf{Defender capabilities.} We assume a defender who may inspect a Skill's contents before installation, who may configure whatever isolation and permission controls the chosen runtime provides, and who may monitor the agent's filesystem and network activity. The defender's inspection is bounded by what a Skill's contents reveal, which does not extend to the full set of behaviors those contents will elicit across the space of possible user requests.

\textbf{Scope.} Our analysis covers threats that arise from the Agent Skills framework itself: from the format of a Skill package, from the process by which Skills are distributed and installed, and from the manner in which an activated Skill's contents are loaded and acted upon. Attacks on the language model itself, including jailbreaks, training-data poisoning, and weight manipulation, fall outside this scope, as do vulnerabilities in the host operating system, implementation defects in an agent runtime, and compromise of the model provider's infrastructure. We include the case in which a Skill causes such a vulnerability to be reached, but not the vulnerability itself.

\subsection{Lifecycle Attack Surface}\label{sec:lifecycle:phases}

\begin{figure}[t]
  \centering
  \includegraphics[width=\textwidth]{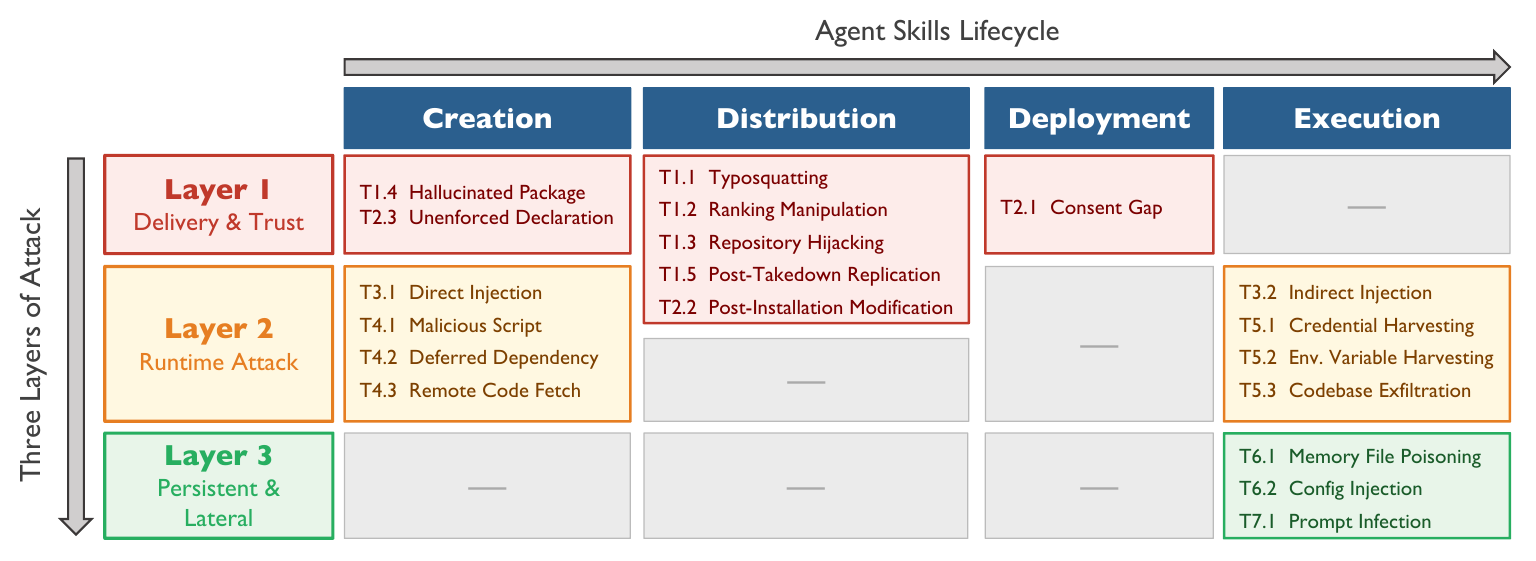}
  \caption{The Agent Skills lifecycle and threat taxonomy. The horizontal axis represents the four lifecycle phases. The vertical axis organizes threats into three attack layers. Scenarios that Table~\ref{tab:taxonomy} maps to two phases are placed here at the earlier of the two, the phase in which they originate. Empty cells indicate that no threat in that layer originates at that phase.}
  \label{fig:lifecycle}
  \Description{The Agent Skills lifecycle and threat taxonomy. The horizontal axis represents the four lifecycle phases. The vertical axis organizes threats into three attack layers. Scenarios that Table~\ref{tab:taxonomy} maps to two phases are placed here at the earlier of the two, the phase in which they originate. Empty cells indicate that no threat in that layer originates at that phase.}
\end{figure}

An Agent Skill traverses four phases from authorship to runtime. In the \emph{Creation} phase, a Skill author composes the \texttt{SKILL.md} file and any bundled scripts or assets. In the \emph{Distribution} phase, the packaged Skill is published to a marketplace or repository and made available for installation. In the \emph{Deployment} phase, a user installs the Skill into their agent environment and the framework grants it authority. In the \emph{Execution} phase, the agent activates the Skill at runtime and acts on its instructions. Figure~\ref{fig:lifecycle} illustrates the lifecycle and the primary attack entry points at each phase.

The four phases form a cycle rather than a single pass. A published Skill may be revised, and each revision re-enters Distribution while the installed base remains at Deployment, carrying an approval that was granted against the earlier content. A Skill removed from a marketplace likewise does not leave the cycle, because copies persist in downstream repositories and on the machines of users who installed it. Threats that exploit this re-entry are distinguished in Section~\ref{sec:taxonomy} from those that arise on a Skill's first pass.

\textbf{Creation.} At the Creation phase, the Skill author has unrestricted control over both the \texttt{SKILL.md} instructions body and any bundled executable scripts. A malicious author may craft a \texttt{description} that appears benign while embedding adversarial directives in the instructions body, where they will carry the same standing as the Skill's legitimate content once it is activated. The gap between declared and actual behavior is not detectable at authorship time, because establishing that an instructions body is confined to its declared purpose would require a specification of that purpose precise enough to check against, and the \texttt{description} field provides no such specification. Supply chain attackers who compromise a legitimate author's credentials or repository inherit the same degree of control without requiring any authorship access.

\textbf{Distribution.} At the Distribution phase, the Skill is published to a marketplace or shared repository. The absence of mandatory security review in the major Agent Skills marketplaces means that malicious Skills may reach large user populations without any vetting checkpoint~\cite{snyk2026toxicskills}. Distribution-phase attacks target the two mechanisms by which a Skill is selected. The first is human selection, which an attacker influences by registering Skill names that closely resemble popular legitimate ones, or by inflating the download counts and ratings that users treat as quality signals. The second is the agent's relevance function $r(q, m_i)$, which an attacker influences by crafting a \texttt{description} that maximizes selection probability for high-value task contexts. Because $r$ operates over author-supplied natural language rather than over verified identifiers, both operands of the selection decision are under the attacker's control.

\textbf{Deployment.} At the Deployment phase, the user installs the Skill and the framework grants it the authority described in Section~\ref{sec:architecture:trustmodel}. The consent gap is structurally located here. The object of the user's approval is the declared $\mathcal{D}_i$ together with whatever content the user reviews at installation time, while the scope actually granted is $\mathcal{A}$, persistent across sessions and unaffected by subsequent changes to the Skill's content. A Skill whose declared scope is narrow at install time may later direct the agent to perform actions drawn from anywhere in $\mathcal{A}$, either through content modified after installation or through instructions that were present but not scrutinized.

\textbf{Execution.} At the Execution phase, the agent reads the activated Skill's \texttt{SKILL.md} into its context window and acts on its instructions. This phase is the primary site of prompt injection and runtime privilege abuse. Any adversarial directive the instructions body contains is interpreted with the same standing as legitimate Skill instructions, whether it was introduced at Creation, injected via a compromised supplementary file, or retrieved from a network resource that the Skill directs the agent to fetch. The progressive disclosure model extends this surface beyond \texttt{SKILL.md} itself, because supplementary files and script outputs that enter the context at Level 3 may carry adversarial content of their own. Script execution presents a distinct risk. Since script source never enters the context window and only its output is returned, a malicious script may perform arbitrary filesystem or network operations that the agent cannot observe before acting on the result.

The four phases and their associated attack surfaces define the scope of the threat taxonomy developed in the following section. Threats are organized by their primary phase of origin, though many span multiple phases. A supply chain compromise introduced at Distribution may manifest as a prompt injection at Execution, with the Deployment phase's consent gap preventing the user from recognizing that the authority being exercised was never legitimately granted.

\section{Threat Taxonomy}\label{sec:taxonomy}

\subsection{Construction Methodology}\label{sec:taxonomy:method}

We construct the taxonomy by combining two sources of evidence, a review of public reporting on attacks against Agent Skills and analogous extension mechanisms, and a structural analysis of the framework described in Section~\ref{sec:architecture}.

\textbf{Source collection.} We searched four classes of source. For peer-reviewed and preprint research we queried the ACM Digital Library, IEEE Xplore, USENIX proceedings, and arXiv. For vulnerability records we queried the National Vulnerability Database. For vendor and practitioner reporting we collected security advisories and technical write-ups from security firms and from the platform operators themselves. For ecosystem evidence we collected marketplace incident announcements and takedown notices. Queries combined the terms \emph{agent skill}, \emph{SKILL.md}, \emph{skill injection}, \emph{skill marketplace}, and \emph{agent extension} with \emph{security}, \emph{attack}, \emph{vulnerability}, and \emph{supply chain}. The search window opens in October 2025, when the format was introduced. We additionally followed citations from every included source in both directions.

\textbf{Selection.} We retained a source if it described a concrete attack technique, reported a measured vulnerability prevalence, or disclosed an incident affecting Agent Skills or a mechanism whose relevant architectural properties Agent Skills share. We excluded sources that discussed agent security without reference to a capability extension mechanism, and sources whose only technical content was reproduced from a source already retained. Where several outlets reported the same incident, we retained the primary technical account and used secondary reporting only where it supplied details the primary account omitted.

\textbf{Structural analysis.} We enumerated the attack surfaces implied by the architectural properties in Section~\ref{sec:architecture}, asking for each what an adversary could do that the property fails to prevent, without regard to whether such an attack had been reported. We then merged the two sets. Scenarios derived structurally but not yet observed were retained when they follow from a documented architectural property, and are marked as such in Table~\ref{tab:taxonomy} so that architectural inference is not presented as measured observation. Scenarios drawn from public reporting were assigned to the architectural property that makes them possible. Scenarios are individuated by attack vector rather than by consequence, so that a single technique used for two ends appears once.

\subsection{Structure of the Taxonomy}\label{sec:taxonomy:structure}

Table~\ref{tab:taxonomy} organizes nineteen scenarios into seven categories across three attack layers. The first layer covers how a malicious Skill reaches users and acquires authority, through supply chain compromise and consent abuse. The second layer covers the attacks an activated Skill can mount, through prompt injection, code execution, and data exfiltration. The third layer covers how the effects of a compromise outlast the session or cross the agent boundary, through persistence and multi-agent propagation.

Each scenario carries three annotations. \emph{Origin} records whether the scenario follows from the specification, from the latitude the specification leaves to individual runtimes, or from how the ecosystem distributes Skills, using the same division as Table~\ref{tab:comparison}. \emph{Evidence} records the strength of the public record supporting the scenario. \emph{Feasibility} records whether mitigation is available without changing the specification, requires a specification change, or remains open. The three annotations together indicate where the framework is most exposed and where remediation is within reach, and Section~\ref{sec:discussion} develops the feasibility classification in detail.

\begingroup
\footnotesize
\setlength{\tabcolsep}{4pt}
\begin{longtable}{
  >{\raggedright\arraybackslash}m{\dimexpr 0.06\linewidth-0.84\tabcolsep\relax}
  >{\raggedright\arraybackslash}m{\dimexpr 0.205\linewidth-2.87\tabcolsep\relax}
  >{\raggedright\arraybackslash}m{\dimexpr 0.405\linewidth-5.67\tabcolsep\relax}
  >{\centering\arraybackslash}m{\dimexpr 0.1\linewidth-1.4\tabcolsep\relax}
  >{\centering\arraybackslash}m{\dimexpr 0.08\linewidth-1.12\tabcolsep\relax}
  >{\centering\arraybackslash}m{\dimexpr 0.075\linewidth-1.05\tabcolsep\relax}
  >{\centering\arraybackslash}m{\dimexpr 0.075\linewidth-1.05\tabcolsep\relax}
}
\caption{Threat taxonomy for Agent Skills. \textbf{Ph} gives the lifecycle phase of origin: Cr = Creation, Di = Distribution, De = Deployment, Ex = Execution. \textbf{Or} gives the origin of the exposure: Sp = specification, Im = implementation latitude, Ec = ecosystem distribution. \textbf{Ev} gives the strength of the public evidence: M = multiple independent measurements or campaigns, D = a single documented incident or demonstration, A = architectural inference with no public instance. \textbf{Fe} gives the mitigation feasibility: I = incremental, deployable without specification change, S = requires a specification change, O = open problem.}
\label{tab:taxonomy} \\

\toprule
\textbf{ID} & \textbf{Scenario} & \textbf{Description} & \textbf{Ph} & \textbf{Or} & \textbf{Ev} & \textbf{Fe} \\
\midrule
\endfirsthead

\multicolumn{7}{c}{\tablename~\thetable{} (continued)} \\
\toprule
\textbf{ID} & \textbf{Scenario} & \textbf{Description} & \textbf{Ph} & \textbf{Or} & \textbf{Ev} & \textbf{Fe} \\
\midrule
\endhead

\midrule
\multicolumn{7}{r}{\textit{Continued on next page}} \\
\endfoot

\bottomrule
\endlastfoot

\rowcolor{groupbg}
\multicolumn{7}{l}{\textbf{Layer 1: Delivery and Trust Establishment}} \\

\rowcolor{groupbg}
\multicolumn{7}{l}{\textit{T1: Supply Chain Compromise}} \\
T1.1 & Typosquatting & Malicious Skill registered under a name resembling a popular legitimate Skill to deceive users into installation & Di & Ec & M & I \\
\rowcolor{rowgray}
T1.2 & Ranking Manipulation & Attacker inflates download counts and ratings to position a malicious Skill above legitimate alternatives & Di & Ec & M & I \\
T1.3 & Repository Hijacking & Attacker gains control of a legitimate Skill repository through account takeover & Di & Ec & D & I \\
\rowcolor{rowgray}
T1.4 & Hallucinated Package & Skill references packages that do not exist, which attackers later claim on public registries to achieve code execution & Cr/Di & Ec & D & I \\
T1.5 & Post-Takedown Replication & Copies of a removed Skill persist in downstream repositories, aggregators, and installed environments after marketplace removal & Di & Ec & D & O \\

\midrule
\rowcolor{groupbg}
\multicolumn{7}{l}{\textit{T2: Consent Abuse}} \\
\rowcolor{rowgray}
T2.1 & Consent Gap & Authority granted at installation spans the agent's full action space rather than the Skill's declared scope & De/Ex & Sp & M & S \\
T2.2 & Post-Installation Modification & Skill content is modified after installation, inheriting the original trust grant without re-approval & Di/Ex & Sp & D & S \\
\rowcolor{rowgray}
T2.3 & Unenforced Declaration & Author declares a restricted tool set that some runtimes honor and others silently discard, so the declaration misleads rather than constrains & Cr/Ex & Sp & A & S \\

\midrule
\rowcolor{groupbg}
\multicolumn{7}{l}{\textbf{Layer 2: Runtime Attack}} \\

\rowcolor{groupbg}
\multicolumn{7}{l}{\textit{T3: Prompt Injection}} \\
T3.1 & Direct Injection & Adversarial instructions authored into \texttt{SKILL.md}, in the instructions body or in the frontmatter that loads before any activation & Cr/Ex & Sp & M & O \\
\rowcolor{rowgray}
T3.2 & Indirect Injection & Skill retrieves external content containing adversarial instructions, which enters the same context as the Skill's own directives & Ex & Sp & M & I \\

\midrule
\rowcolor{groupbg}
\multicolumn{7}{l}{\textit{T4: Code Execution}} \\
T4.1 & Malicious Script & Bundled script executes arbitrary system commands, including ransomware deployment or credential theft & Cr/Ex & Im & M & I \\
\rowcolor{rowgray}
T4.2 & Deferred Dependency & Script declares unpinned dependencies that the attacker later replaces with malicious versions on public registries & Cr/Ex & Ec & D & I \\
T4.3 & Remote Code Fetch & Instructions direct the agent to fetch and execute code from an attacker-controlled URL at runtime, bypassing installation-time review & Cr/Ex & Sp & D & S \\

\midrule
\rowcolor{groupbg}
\multicolumn{7}{l}{\textit{T5: Data Exfiltration}} \\
\rowcolor{rowgray}
T5.1 & Credential Harvesting & Skill directs the agent to read API keys, SSH keys, browser credentials, and wallet data, then transmit them externally & Ex & Im & M & S \\
T5.2 & Environment Variable Harvesting & Scripts access and exfiltrate environment variables containing secrets from the agent's runtime & Ex & Im & A & S \\
\rowcolor{rowgray}
T5.3 & Codebase Exfiltration & Skill reads and transmits the project codebase with no visible indication in agent output or audit logs & Ex & Im & D & S \\

\midrule
\rowcolor{groupbg}
\multicolumn{7}{l}{\textbf{Layer 3: Persistent and Lateral Impact}} \\

\rowcolor{groupbg}
\multicolumn{7}{l}{\textit{T6: Persistence}} \\
T6.1 & Memory File Poisoning & Skill writes adversarial content into persistent instruction files, altering agent behavior in sessions where the Skill is never activated & Ex & Im & D & I \\
\rowcolor{rowgray}
T6.2 & Config Injection & Skill modifies agent configuration files to establish persistent backdoors or pre-authorize dangerous operations & Ex & Im & M & I \\

\midrule
\rowcolor{groupbg}
\multicolumn{7}{l}{\textit{T7: Multi-Agent Propagation}} \\
T7.1 & Prompt Infection & A Skill-controlled agent propagates adversarial instructions to downstream agents in a multi-agent pipeline & Ex & Sp & A & S \\

\end{longtable}
\endgroup

\subsection{T1: Supply Chain Compromise}\label{sec:taxonomy:t1}

Supply chain compromise attacks target the distribution of Skills rather than their content. The goal is to position a malicious Skill for installation by users who believe they are installing a legitimate one. The absence of mandatory review in the major marketplaces, combined with the low barrier to publishing, produces conditions analogous to the early years of npm and PyPI, with the difference that a compromised Skill acts inside the user's own environment rather than inside a single application.

\textbf{T1.1 Typosquatting.} An attacker registers a Skill under a name that closely resembles a popular legitimate one. The attack works because neither users nor agents verify publisher identity cryptographically before selection. The ClawHavoc campaign applied this technique at scale, and the permissive upload model of the affected marketplace, which accepted publications from any account older than one week, kept the cost of each attempt near zero~\cite{antiy2026clawhavoc}.

\textbf{T1.2 Ranking Manipulation.} An attacker inflates the download counts, ratings, or endorsement scores of a malicious Skill so that it appears above legitimate alternatives in marketplace listings. Mitiga observed coordinated campaigns in which automated accounts mass-downloaded malicious Skills and submitted positive reviews~\cite{mitiga2026agentskills}, and Authmind recorded the same pattern during ClawHavoc, with one malicious Skill reaching the top position in its category before discovery~\cite{authmind2026agentskills}. The same principle extends to the agent's own selection, since selection at Level 2 is mediated by semantic matching over descriptions rather than by exact name lookup, and a description written to maximize apparent relevance is the model-facing analogue of an inflated rating~\cite{ama2026metadata}. Because both users and automated installation tools prefer highly ranked Skills, manipulation of these signals translates directly into installation volume.

\textbf{T1.3 Repository Hijacking.} An attacker obtains control of a legitimate and previously trusted Skill repository through account takeover, credential theft, or claim of an abandoned namespace, then pushes malicious updates through the original distribution channel. The attack inherits the reputation and the installed base of the compromised Skill. SafeDep identified it as a primary vector for Agent Skills, drawing on the extensive record of analogous attacks in npm and PyPI~\cite{safedep2026agentskills,ladisa2023sok}. Unlike typosquatting, which requires a user to be deceived into a new installation, repository hijacking delivers its payload as a routine update to users who have already extended trust.

\textbf{T1.4 Hallucinated Package.} A Skill's bundled scripts reference package names that do not exist on public registries, a consequence of language models generating plausible but fictitious dependency names. An attacker monitors registries for such names and claims them, so that a malicious package is installed the next time the Skill executes. SafeDep identified this as specific to AI-generated Skill content~\cite{safedep2026agentskills}, and Aikido Security documented Skills whose instructions invoked hallucinated \texttt{npx} commands that attackers subsequently registered~\cite{aikido2026agentskills}. The attack requires no compromise of the author's account, and exploits only the gap between what a model generates and what exists.

\textbf{T1.5 Post-Takedown Replication.} Removal of a malicious Skill from a marketplace does not remove it from the ecosystem. Zenity Labs reported a campaign in which the hosting platforms removed the offending entries and repositories within twelve hours of notification, while copies of the same instructions remained in downstream repositories, in aggregator sites that had mirrored the listing, and in the environments of users who had already installed it~\cite{zenity2026skillsupplychain}. The same campaign distributed a binary-identical installation file across seven separate locations, so that removal of any one left the others reachable. Because a Skill is a directory of plain files rather than a package fetched from a single authoritative source at each invocation, no mechanism in the framework propagates a revocation to installed copies. This scenario is the reason we treat the lifecycle in Section~\ref{sec:lifecycle:phases} as a cycle rather than a single pass.

\subsection{T2: Consent Abuse}\label{sec:taxonomy:t2}

Consent abuse attacks exploit the trust model rather than introducing malicious content. The attack surface is the gap between what a user understands themselves to be authorizing and what the framework grants. Three scenarios exploit this gap through different mechanisms.

\textbf{T2.1 Consent Gap.} Using the notation of Section~\ref{sec:architecture:loading}, installation grants a Skill the full action space $\mathcal{A}$, whereas the user's understanding is bounded by the declared scope $\mathcal{D}_i$ and by whatever content they inspect at installation time. The difference $\mathcal{A} \setminus \mathcal{D}_i$ is available to the Skill in every subsequent session without further confirmation. Cherny documented the consequence in the MedusaLocker incident, where a Skill presented as a GIF converter was approved by users who had no reason to anticipate that their approval covered persistent filesystem writes, outbound network connections, and executable downloads~\cite{cherny2025medusalocker}. Liu et al.\ identified exploitation of this gap as the mechanism underlying the privilege escalation findings in their audit~\cite{liu2026agent}, and the OWASP catalogue for agentic Skills classifies the pattern as identity and privilege abuse~\cite{owasp2026skills}.

\textbf{T2.2 Post-Installation Modification.} After installation has established a trust grant, the author or an attacker who has taken over the distribution channel modifies the Skill's content. Because the grant is bound to the Skill's identity rather than to a committed version of its content, the modified Skill inherits the original approval without re-authorization. Snyk documented Skills whose initial content was benign and which were structured to receive a malicious payload once an installed base had accumulated~\cite{snyk2026toxicskills}. The scenario is the Skills analogue of the rug pull documented by Hou et al.\ for MCP~\cite{hou2025mcp}, and is more severe here because no mechanism notifies a user that installed content has changed.

\textbf{T2.3 Unenforced Declaration.} An author restricts a Skill using the optional frontmatter fields described in Section~\ref{sec:architecture:package}, and a user installs it in reliance on that restriction. Because the specification defines what an author may declare without defining what a runtime must do with the declaration, whether the restriction takes effect depends on the runtime. Claude Code applies \texttt{allowed-tools} when authorizing tool use, whereas the Agent SDK does not read the field~\cite{anthropic2026skillsbestpractices}. A restriction that is honored in the environment where a Skill is reviewed and discarded in the environment where it is deployed leaves the user worse informed than an absent declaration would, because it supplies a reason for confidence that does not hold. The scenario follows from the documented inconsistency between runtimes.

\subsection{T3: Prompt Injection}\label{sec:taxonomy:t3}

Prompt injection attacks introduce adversarial instructions into the context on which the agent acts. Unlike injection attacks against conventional software, which exploit a parser's failure to separate code from data, prompt injection against Agent Skills exploits the fact that no such separation is defined. Two scenarios follow, distinguished by whether the adversarial content is authored into the Skill or retrieved by it at runtime.

\textbf{T3.1 Direct Injection.} The Skill author places adversarial instructions in \texttt{SKILL.md} itself. The attack exploits the absence of a contract between the frontmatter and the instructions body, so that the \texttt{description} may declare one purpose while the file encodes another. Both parts of the file serve as carriers, and they differ in reach. Instructions placed in the body enter the context at Level 2, when the agent activates the Skill in response to a request it judges relevant. The frontmatter enters earlier and unconditionally: as described in Section~\ref{sec:architecture:loading}, the metadata pair $m_i$ of every installed Skill is injected into the system prompt at Level 1, before any user request and without regard to whether the Skill is ever activated. Content placed there acts on the agent by installation alone, and it also inverts the usual review assumption, since a reviewer who scrutinizes the instructions body and skims the frontmatter has inspected the less consequential part of the file.

Schmotz et al.\ showed that the attack requires no technical sophistication, with three lines of Markdown sufficient to redirect agent behavior~\cite{schmotz2025trivial,schmotz2026skillinject}, and subsequent work has automated the generation of poisoned Skills through instruction rewriting and helper scripts~\cite{jia2026skillject,badskill2026}. Work on metadata-level attacks against tool selection has shown that descriptions optimized for the model rather than for the human reader reliably capture invocation decisions~\cite{ama2026metadata}. The audit by Liu et al.\ of 42,447 Skills found injection patterns to be the most prevalent vulnerability class~\cite{liu2026agent}. Detection is difficult for a reason that admits no incremental fix. Distinguishing an injected directive from an intended one requires knowing what the Skill is supposed to do, and the \texttt{description} field is too weak a specification to serve as that reference.

\textbf{T3.2 Indirect Injection.} The Skill itself may be legitimate, but it retrieves external content that an attacker has crafted to contain adversarial instructions. Web pages, documents, API responses, and repository issues all serve as carriers. When such content enters the context at Level 3, it sits alongside the Skill's own directives with nothing marking the difference. Schmotz et al.\ demonstrated the attack in the Agent Skills setting, showing that retrieved content can redirect an agent to act outside the retrieving Skill's declared scope~\cite{schmotz2026skillinject}. Greshake et al.\ introduced the threat class for LLM-integrated applications generally~\cite{greshake2023not}, and the Red Hat security team identified it as unresolved for Skill deployments~\cite{redhat2026agentskills}. The exposure is widespread rather than incidental. A survey of 142,836 published Skills found that 17,822 of them, accounting for 6.7 million installations, reference untrusted external instruction sources~\cite{air2026circus}.

\subsection{T4: Code Execution}\label{sec:taxonomy:t4}

Code execution attacks exploit the co-location of natural language instructions and executable scripts within one package. Where prompt injection operates on the instructions the agent reads, these attacks operate on code that runs outside the context window, so that the behavior is unavailable to the model before it acts on the result.

\textbf{T4.1 Malicious Script.} An author bundles a script that executes arbitrary system commands when the agent invokes it. Because only the script's output returns to the context, the model cannot inspect what the script did before acting on that output. Cato CTRL researchers demonstrated the attack concretely with a Skill advertised as a GIF converter, which downloaded and executed MedusaLocker ransomware while returning benign output to the agent~\cite{cherny2025medusalocker}. The ClawHavoc campaign showed the same pattern at scale, with malicious scripts deploying macOS and Windows infostealers across the compromised Skills~\cite{antiy2026clawhavoc,liu2026malicious}. The severity of this scenario depends on the runtime, since a Skill executing under filesystem and network isolation reaches far less than one executing at the user's privilege level.

\textbf{T4.2 Deferred Dependency.} A bundled script declares dependencies without pinning them to specific versions. The script is benign when reviewed, and the attacker publishes a malicious version of a referenced package afterwards. On the next execution the runtime resolves and installs the current version, which now carries the payload. SafeDep demonstrated this using Python's PEP 723 inline script metadata, showing that \texttt{uv run} resolves and installs unpinned dependencies at execution time without notifying the user~\cite{safedep2026agentskills}. The attack defeats installation-time review by construction, because at review time no malicious code exists in the supply chain.

\textbf{T4.3 Remote Code Fetch.} The instructions body directs the agent to download and execute code from an attacker-controlled URL. Where T4.2 exploits a package registry, this scenario delivers the payload over HTTP directly. Snyk documented Skills instructing agents to pipe the contents of a remote URL into a shell~\cite{snyk2026toxicskills}. The file under review contains a fetch instruction rather than a payload, and the fetched content can be changed at any time, so a Skill can serve benign content during a review period and malicious content afterwards.

\subsection{T5: Data Exfiltration}\label{sec:taxonomy:t5}

Data exfiltration attacks read sensitive data from the local environment and transmit it to attacker-controlled infrastructure. Skills are well suited to this class because the progressive disclosure model allows a Skill to read files on demand without per-file confirmation, and because in unisolated deployments the agent's filesystem and network reach is the user's own. Three targets define the scenarios.

\textbf{T5.1 Credential Harvesting.} The Skill directs the agent to read authentication material from predictable filesystem locations and to transmit it. The ClawHavoc campaign harvested API keys, SSH private keys, browser-stored passwords, and cryptocurrency wallet data across dozens of categories~\cite{antiy2026clawhavoc}. Cisco's AI Defense team found that the most popular community Skill on a major marketplace was functional malware that exfiltrated credentials while performing its advertised task~\cite{cisco2026skillscanner}. The July 2026 campaign reported by Zenity Labs extended the target set to the credentials that characterize developer and continuous integration environments, including cloud provider tokens, package manager credentials, container and orchestration configuration, and project environment files~\cite{zenity2026skillsupplychain}.

\textbf{T5.2 Environment Variable Harvesting.} A bundled script reads environment variables from the agent's runtime, which commonly hold API keys, database connection strings, and cloud credentials injected at deployment. SafeDep identified this as a distinct vector, noting that environment variables aggregate secrets from multiple services into one programmatically accessible location~\cite{safedep2026agentskills}. Unlike credential files at known paths, environment variables leave no filesystem access trace, so exfiltration through this channel is harder to detect. It is separated from T5.1 because the detection surface differs.

\textbf{T5.3 Codebase Exfiltration.} The Skill reads and transmits a project codebase. Mitiga demonstrated the attack in their Breaking Skills series, exfiltrating a complete codebase in four user interactions and leaving the audit log empty~\cite{mitiga2026agentskills}. The consequence is most serious in enterprise development environments, where codebases carry proprietary algorithms, unreleased functionality, infrastructure configuration, and embedded secrets. Because the transfer uses the same network channels the agent uses legitimately, distinguishing it from ordinary operation requires monitoring built for the purpose.

\subsection{T6: Persistence}\label{sec:taxonomy:t6}

Persistence attacks write to durable locations in the agent's environment, so that the effect of a compromise survives the session and, in most cases, the removal of the Skill that caused it. Two mechanisms enable persistence.

\textbf{T6.1 Memory File Poisoning.} The Skill writes adversarial content into the instruction files that the agent reads at startup or consults during a task, such as the \texttt{AGENTS.md} and \texttt{CLAUDE.md} files described in Section~\ref{sec:architecture:trustmodel}. These files are loaded as standing guidance, and the agent has no basis for distinguishing an entry a user wrote from one a Skill inserted. Snyk identified the vector and recommended that users inspect these files after installing Skills from untrusted sources~\cite{snyk2026toxicskills}. Chen et al.\ established the broader class against retrieval-based agents, showing that adversarial content written into an agent's knowledge base redirects behavior across arbitrarily many later sessions~\cite{chen2024agentpoison}. The scenario is the reason that removing a malicious Skill is not sufficient remediation.

\textbf{T6.2 Config Injection.} The Skill modifies agent configuration files to establish backdoors, pre-authorize dangerous operations, or redirect API traffic. Check Point Research documented two instances against Claude Code. CVE-2025-59536 showed that a malicious repository could inject hook configurations that execute shell commands at agent initialization, before any trust dialog is presented~\cite{checkpointresearch2026}. CVE-2026-21852 showed that setting the API base URL in project configuration routes all traffic, including authentication tokens, to an attacker-controlled endpoint~\cite{checkpointresearch2026}. Configuration changes affect every subsequent operation of the agent rather than only those involving the Skill that made them.

\subsection{T7: Multi-Agent Propagation}\label{sec:taxonomy:t7}

Multi-agent propagation attacks use the trust relationships between agents in an orchestrated pipeline to extend a Skill-level compromise across the system. As agents are increasingly deployed in configurations where one delegates to others, a compromised agent becomes a vector for agents that have installed no malicious Skill of their own.

\textbf{T7.1 Prompt Infection.} A Skill-controlled agent embeds adversarial instructions in the messages it sends downstream. The receiving agents treat the inter-agent channel as trusted and act on the instructions with whatever authority their own trust model affords. Lee and Tiwari formalized the class, showing that adversarial instructions can replicate across interconnected agents, each infected agent infecting those it contacts~\cite{lee2024prompt}. In the Agent Skills setting the consequence is that a single installation by one agent in a pipeline can affect agents operating at higher privilege, including orchestrators with access to production infrastructure. Current agent frameworks provide no standardized trust model for inter-agent communication against which such a boundary could be enforced~\cite{yu2025survey}.

\section{Real-World Incidents}
\label{sec:incidents}

This section examines seven documented incidents in the Agent Skills ecosystem and maps each to the taxonomy of Section~\ref{sec:taxonomy}.

\textbf{Selection.} We include an incident when it satisfies four conditions. It must be publicly disclosed with a technical account from the party that discovered it, rather than reported only at second hand. It must involve Agent Skills specifically, rather than an agent vulnerability that Skills merely illustrate. It must be distinguishable in root cause from the incidents already included. And its account must be detailed enough to determine which taxonomy scenarios it exercises. The incidents fall into three groups. The first covers confirmed attacks against individual users. The second covers ecosystem-scale supply chain campaigns and platform-level vulnerabilities. The third covers researcher-conducted attacks against live deployments. Table~\ref{tab:incidents} summarizes the mappings.

\begin{table}[t]
\centering
\small
\setlength{\tabcolsep}{4pt}
\renewcommand{\arraystretch}{1.4}
\caption{Documented Agent Skills security incidents and their taxonomy mappings.}
\label{tab:incidents}
\begin{tabular}{
  >{\raggedright\arraybackslash}m{\dimexpr 0.27\linewidth-1.62\tabcolsep\relax}
  >{\raggedright\arraybackslash}m{\dimexpr 0.23\linewidth-1.38\tabcolsep\relax}
  >{\raggedright\arraybackslash}m{\dimexpr 0.5\linewidth-3.0\tabcolsep\relax}
}
\toprule
\textbf{Incident} & \textbf{Mapping} & \textbf{Key Impact} \\
\midrule
MedusaLocker Skill & T4.1, T2.1 & Ransomware executed via bundled script, consent gap exploited \\
\rowcolor{rowgray}
ClawHavoc Campaign & T1.1, T1.2, T5.1 & At least 1,184 malicious Skills, credential theft at ecosystem scale \\
CVE-2025-59536 / CVE-2026-21852 & T6.2, T5.1 & Command execution and API key exfiltration via configuration injection \\
\rowcolor{rowgray}
SafeDep PEP 723 & T4.2 & Deferred dependency attack via unpinned script dependencies \\
Mitiga Silent Egress & T5.3, T1.2 & Project codebase exfiltrated in four user interactions, audit log empty \\
\rowcolor{rowgray}
Zenity Skill Supply Chain & T1.1, T1.2, T1.5, T5.1 & 1.7 million installations, developer and CI credentials harvested, copies persisted after takedown \\
Air Security Trusted Channel & T3.2, T2.2 & Payload served from external documentation URL after review, 26,000 agents affected, cleared by every scanner \\
\bottomrule
\end{tabular}
\end{table}

\subsection{MedusaLocker Ransomware Skill}\label{sec:incidents:medusa}

\begin{figure}[t]
  \centering
  \includegraphics[width=\columnwidth]{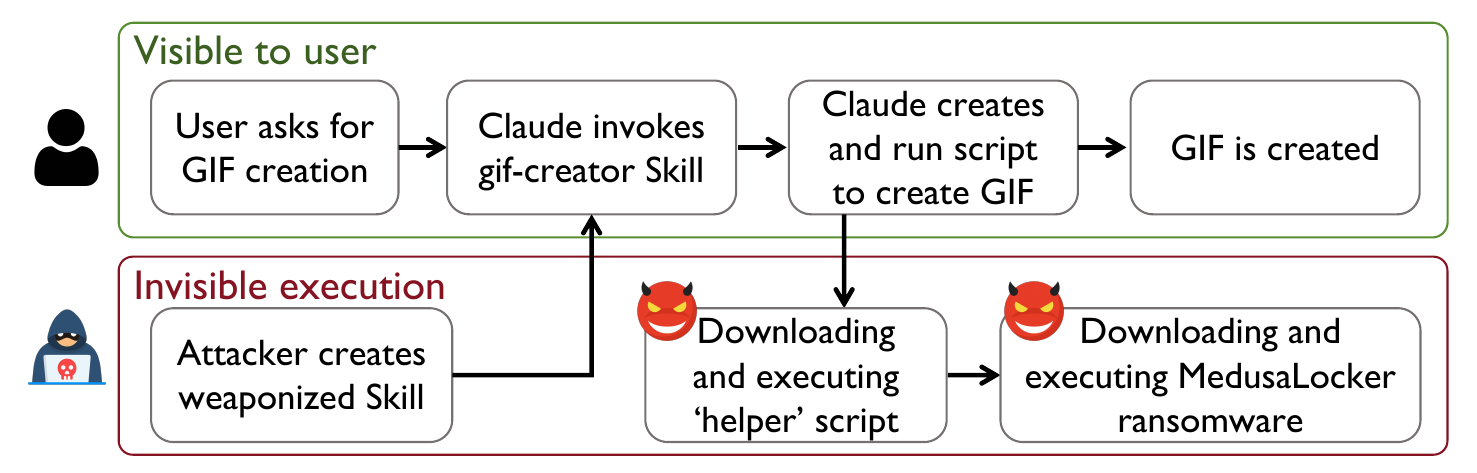}
  \caption{The MedusaLocker Ransomware attack. The user-visible layer shows normal GIF creation behavior, while the invisible execution layer downloads and executes MedusaLocker ransomware through the bundled helper script.}
  \label{fig:medusa}
  \Description{The MedusaLocker Ransomware attack. The user-visible layer shows normal GIF creation behavior, while the invisible execution layer downloads and executes MedusaLocker ransomware through the bundled helper script.}
\end{figure}

In December 2025, researchers at Cato CTRL demonstrated the first publicly documented execution of ransomware through a weaponized Agent Skill~\cite{cherny2025medusalocker}. The Skill was presented as a GIF image conversion utility, a plausible and low-risk capability that required no elevated permissions in its declared \texttt{description}. Once installed and activated, the bundled script downloaded and executed MedusaLocker ransomware, encrypting the user's filesystem while returning benign output to the agent. Figure~\ref{fig:medusa} illustrates the two-layer execution model.

The incident exercises two scenarios. As an instance of \textbf{T4.1}, it turns on the loading model described in Section~\ref{sec:architecture:loading}. Because a bundled script's source never enters the context window, the agent had no basis for judging what the script would do before acting on its output, and the payload was delivered and executed entirely outside the agent's reasoning. As an instance of \textbf{T2.1}, the authority the user granted was $\mathcal{A}$ while the authority they understood themselves to be granting was bounded by a GIF converter's declared scope. Nothing in the approval distinguished filesystem encryption from image conversion.

The incident shows that the consent gap does not depend on a Skill's declared purpose. Any Skill acquires the same authority on installation, so the plausibility of the cover story bears on whether a user installs the Skill but not on what the Skill can then do. The attack required only the ability to publish a Skill with a credible description and a malicious script.

\subsection{ClawHavoc Campaign}\label{sec:incidents:clawhavoc}

Koi Security began auditing the ClawHub marketplace, the primary distribution channel for the OpenClaw agent ecosystem, in late January 2026, and disclosed a coordinated poisoning campaign on February 1. Antiy CERT extended the analysis and by February 5 attributed at least 1,184 malicious Skills to twelve publisher accounts, of which a single account contributed 677~\cite{antiy2026clawhavoc,liu2026malicious}. Malicious Skills deployed macOS and Windows infostealers that harvested LLM API keys, SSH private keys, browser-stored passwords, and cryptocurrency wallet data across dozens of categories.

The campaign combined three scenarios. Attackers registered Skills under names resembling popular legitimate ones (\textbf{T1.1}), and inflated download counts and ratings so that malicious Skills appeared above legitimate alternatives (\textbf{T1.2}). Once installed, the Skills directed agents to locate and transmit credentials from predictable filesystem locations (\textbf{T5.1}).

Two properties of the ecosystem made the campaign inexpensive. Publication required only a \texttt{SKILL.md} file and an account older than one week, so each attempt cost the attacker nothing beyond the time to prepare a package. And because no marketplace binds a publication to a verified identity, the twelve accounts were not distinguishable in advance from legitimate new publishers. Where provenance cannot be verified, popularity is the only remaining quality signal available to a user, and popularity is precisely what T1.2 manipulates.

\subsection{CVE-2025-59536 and CVE-2026-21852}\label{sec:incidents:cve}

In February 2026, Check Point Research disclosed two vulnerabilities in Claude Code that demonstrated configuration injection as a practical attack vector~\cite{checkpointresearch2026}. CVE-2025-59536 (CVSS 8.7) covered two related flaws. A malicious repository could inject hook configurations into \texttt{.claude/\allowbreak settings.json} that execute shell commands at agent initialization, before any trust dialog appears, and \texttt{.mcp.json} could be configured to auto-approve all MCP servers, triggering execution of attacker-controlled servers on repository load. CVE-2026-21852 (CVSS 5.3) demonstrated that setting \texttt{ANTHROPIC\_\allowbreak BASE\_\allowbreak URL} to an attacker-controlled endpoint in project configuration caused Claude Code to route all API traffic, including authentication tokens, to the attacker before the user was notified.

Both map to \textbf{T6.2}, and CVE-2026-21852 additionally to \textbf{T5.1}. The vulnerabilities were patched in Claude Code versions 1.0.111 and 2.0.65 respectively. These two are the only incidents in our set whose root cause is an implementation defect rather than a property of the specification. The specification does not prescribe where an agent stores configuration or how it decides to trust it, so a runtime that requires explicit confirmation before applying repository-supplied configuration remains conforming. The exposure recurs, since filesystem-based configuration is how every current runtime is built, yet it is remediable within a single implementation in a way that T2.1 and T3.1 are not. Table~\ref{tab:taxonomy} records this distinction in its Origin column.

\subsection{SafeDep PEP 723 Deferred Dependency Attack}\label{sec:incidents:safedep}

In January 2026, SafeDep researchers demonstrated a deferred dependency attack using Python's PEP 723 inline script metadata format~\cite{safedep2026agentskills}. An author publishes a script with a benign implementation that declares a dependency without pinning it to a version. At publication and review time the referenced package either does not exist or contains only benign code. The attacker later publishes a malicious version to PyPI, and on subsequent executions \texttt{uv run} resolves and installs the current version, delivering the payload with no change to the Skill's own files.

The incident maps to \textbf{T4.2}. Its significance is temporal rather than technical. The code that runs when a Skill executes is not fixed at installation, so review at installation establishes nothing about what will run later. The attack is not merely hard to detect at review time, it is undetectable then, because at that moment no malicious artifact exists anywhere in the supply chain.

The incident locates the security boundary. It cannot be drawn around the Skill package, because a package that is inert at rest can be made active by a change elsewhere. Dependency pinning and lockfiles would close this path at almost no cost to the author.

\subsection{Mitiga Silent Codebase Exfiltration}\label{sec:incidents:mitiga}

In February 2026, Mitiga Labs published a Skill to skills.sh, one of the largest public Agent Skills catalogs, framed as a testing utility~\cite{mitiga2026agentskills}. Its download count and reputation score were inflated through coordinated interactions to establish apparent legitimacy. Once installed and activated by a developer, the Skill read and transmitted the entire project codebase to external infrastructure. The exfiltration required four user interactions and left the audit log empty.

The incident maps to \textbf{T5.3} and \textbf{T1.2}. On-demand file reading exists so that a Skill can load only what a task requires, and the same mechanism lets a Skill read whatever it names without per-file confirmation.

The empty audit log is the part that generalizes. Logging what the agent did is not sufficient when the agent did what it was asked, and the harm lies in what was read and where it was sent rather than in any anomalous action. Detection requires a reference for what this Skill's file access and network activity should look like, which returns to the absence of a declared $\mathcal{D}_i$.

\subsection{Zenity Skill Supply Chain Campaign}\label{sec:incidents:zenity}

Throughout July 2026, a malicious Skill distributed through skills.sh accumulated more than 1.7 million installations before it was disrupted in early August~\cite{zenity2026skillsupplychain}. The collection logic targeted developer workstations, continuous integration runners, and agent workspaces, harvesting SSH keys, cloud provider credentials, Git and package manager tokens, container and orchestration configuration, infrastructure-as-code state, and project environment files. The same installation instructions were replicated across seven separate locations, and the campaign paired the Skill with a malicious package published to PyPI.

The incident maps to \textbf{T1.1} and \textbf{T1.2} in its delivery, and to \textbf{T5.1} in its payload. It is the only incident in our set that exercises \textbf{T1.5}. The hosting platforms removed the offending listings and repositories within twelve hours of notification, which by the standards of marketplace response is fast. Copies nonetheless remained in downstream repositories, in aggregators that had mirrored the listing, and in the environments of the users who had already installed it. The incident shows that response time is not the binding constraint. A Skill is a directory of files that a user copies into their environment, not a package resolved from an authoritative source at each invocation, so nothing in the framework carries a revocation to the copies. Removal at Distribution does not reach the installations that Deployment already produced, and those installations continue to Execute.

The campaign also indicates that the threat has not diminished as the ecosystem matured. It occurred five months after ClawHavoc, on a different marketplace, and against a target set specific to environments where credentials are densest.

\subsection{Trusted-Channel Distribution and External Payload}\label{sec:incidents:air}

In June 2026, Air Security researchers built a malicious Skill, distributed it through legitimate channels, and obtained control of more than 26,000 agents belonging to designers, marketers, and their employers~\cite{air2026story}. Distribution proceeded in two stages. The researchers first submitted the Skill as a pull request to an open-source community repository, which was reviewed and merged, and then advertised it on Instagram to an audience of non-technical professionals. The Skill's own files carried no malicious content. Its instructions directed the agent to an external documentation URL under the researchers' control, and the payload was served from there after the Skill had passed review and accumulated installations. Every security scanner applied to the Skill returned no finding.

The incident maps to \textbf{T3.2} and \textbf{T2.2}. As an instance of T3.2, the adversarial instructions never resided in \texttt{SKILL.md} at any point when it could have been inspected. They arrived at Level 3, from a source the Skill was designed to consult, and entered the context alongside the Skill's own directives. As an instance of T2.2, the modification that weaponized the Skill was made to content the Skill referenced rather than to the Skill itself, so it left no trace in the artifact a user or a marketplace would re-examine. The two scenarios compose into a delivery mechanism that neither addresses alone. T3.2 supplies the injection path, and T2.2 supplies the timing, which allows the same Skill to be benign during review and malicious afterwards.

The scale of this exposure is not incidental to this one Skill. The same research group scanned 142,836 published Skills and found that 17,822 of them, accounting for 6.7 million installations, reference at least one external instruction source with no basis for trusting that it will not change~\cite{air2026circus}.

The security boundary that the SafeDep demonstration placed outside the Skill package for dependencies lies outside it for instructions as well. A Skill that fetches its directives at runtime is a pointer rather than a specification, and reviewing a pointer establishes nothing about what it will resolve to. Scanners match patterns, and there was no pattern to match, because at scan time the Skill did what it claimed.

\subsection{Coverage}\label{sec:incidents:coverage}

The seven incidents exercise six of the seven threat categories. T1, T2, T3, T4, T5, and T6 each have at least one documented instance. T7 rests on the formal treatment of Lee and Tiwari~\cite{lee2024prompt} applied to the Skills execution context, with no documented instance in a Skills deployment. At the scenario level, eleven of the nineteen scenarios in Table~\ref{tab:taxonomy} are exercised by an incident above; the Evidence column records which of the remainder rest on a single external demonstration and which on architectural inference alone.

\section{Discussion}\label{sec:discussion}
\subsection{Defense Directions}

Agent Skills security requires defense at the Skill content layer, the distribution layer, the runtime layer, and the trust model layer. We discuss directions for each threat category and identify where architectural reform is necessary. The Feasibility column of Table~\ref{tab:taxonomy} records the resulting classification for each scenario.

\textbf{Against Supply Chain Compromise (T1).} Mandatory security review before marketplace publication is the most direct mitigation for typosquatting and ranking manipulation, and its effectiveness depends on whether review capacity keeps pace with community contribution~\cite{jagpal2015trends}. Cryptographic provenance verification, requiring Skills to be signed by verified publishers and binding namespace ownership to verified identities, would prevent typosquatting and raise the cost of repository hijacking, and is feasible within the existing specification without breaking the authorship model. A curated registry that catalogs, scans, and signs Skills before distribution demonstrates that this is deployable today~\cite{nvidia2026verified}, though as an additional channel rather than a requirement of the ecosystem. Against ranking manipulation, reputation systems that weight verified installation outcomes rather than raw download counts, and that implement anomaly detection for coordinated inflation patterns, would reduce the effectiveness of bot-driven campaigns. This reform requires marketplace-side changes only and imposes no burden on Skill authors. Against hallucinated packages, static analysis tools that verify the existence of all referenced packages at publication time are implementable today using existing package registry APIs, though they cannot prevent attackers from subsequently claiming and poisoning a previously non-existent package name, a residual risk that dependency pinning at authorship time would close. Post-takedown replication requires a different mechanism. Once a Skill has been copied into a user's environment, nothing in the framework carries a revocation to that copy, and closing this path calls for a revocation channel that the specification does not define.

\textbf{Against Consent Abuse (T2).} The consent gap follows from the trust model, and closing it means changing that model. Fine-grained, per-capability permission grants, analogous to the Android permission model, would allow users to authorize specific action classes at installation time rather than granting undifferentiated authority over the full action space. Version-bound trust, in which the trust grant is cryptographically bound to a specific content hash of the Skill at installation time, would prevent post-installation modification from inheriting the original approval. Both reforms require changes to the Agent Skills specification and the agent runtimes that implement it. Version-bound trust also introduces a usability tension, since it would require re-approval on every Skill update, which conflicts with the seamless update model that gives Agent Skills their operational appeal. A practical compromise is delta-based consent, in which re-approval is triggered only when a content diff exceeds a configurable behavioral threshold. Interim mitigations deployable without specification changes include user-facing changelogs that surface content modifications before the next Skill activation, and re-approval prompts triggered by content changes above a configurable threshold. Content that a Skill references rather than contains falls outside all of these, as in Section~\ref{sec:incidents:air}, where a hash over the Skill's own files would have remained unchanged throughout the attack.

\textbf{Against Prompt Injection (T3).} Defending against direct injection requires the ability to distinguish adversarial instructions from legitimate Skill behavior, which is difficult when both occupy the same natural language medium. Structured query formats such as StruQ~\cite{chen2025struq} and privilege-based instruction hierarchies~\cite{wallace2024instruction} offer partial mitigation by enforcing separation between trusted and untrusted content layers, but these defenses are architecturally inapplicable to Agent Skills, because malicious \texttt{SKILL.md} content already occupies the layer such schemes treat as trusted. Indirect injection is more tractable. Defenses that treat all externally retrieved content as untrusted and apply sanitization before it enters the context window can reduce the attack surface, with manageable overhead for most Skill use cases. Neither approach provides a complete defense against direct injection without a formal behavioral specification of intended Skill conduct, a specification that does not exist and that the natural language instruction model makes difficult to define.

\textbf{Against Code Execution (T4).} Sandboxing confines what a malicious script can reach, and how far it can be applied depends on what the Skill legitimately needs. Executing bundled scripts in container-based or WebAssembly-based isolated environments with restricted capabilities would confine the blast radius of a malicious script. This mitigation is incompatible with Skills whose legitimate function requires broad filesystem access, such as document processing Skills that read and write arbitrary project files, or unrestricted network access. A capability-tiered sandboxing model, in which the isolation level is determined by the Skill's declared capability scope, would preserve utility for legitimate Skills while restricting undeclared access. The specification supplies a partial basis for this in the optional frontmatter fields described in Section~\ref{sec:architecture:package}, but it attaches no enforcement semantics to them, so a runtime cannot treat a declaration as a constraint. Against deferred dependency attacks, version pinning and dependency lockfiles impose minimal authorship overhead and are deployable within the existing specification, and mandating them would close this attack vector without breaking any legitimate use case. Against remote code fetch, outbound network filtering that blocks connections to domains not declared by the Skill is feasible once a declaration binds, and imposes no overhead on Skills that do not require external network access.

\textbf{Against Data Exfiltration (T5).} Purpose-built behavioral monitoring that observes agent network activity and flags transfers inconsistent with the active Skill's declared purpose is the primary defense direction, but its practical effectiveness is limited by the absence of a formal behavioral specification. Without knowing what network behavior a Skill is supposed to exhibit, distinguishing exfiltration from legitimate data transfer requires heuristics that will generate false positives for Skills with broad network permissions. Capability-based permission systems that require Skills to explicitly declare the files they intend to read and the network endpoints they intend to contact would limit the exfiltration surface and make behavioral monitoring tractable. This reform requires specification changes but does not break existing Skills whose declared scope matches their actual behavior. The primary cost is authorship overhead, since Skill authors would need to enumerate capability claims that are currently implicit.

\textbf{Against Persistence (T6).} Integrity monitoring of agent memory and configuration files, detecting unauthorized modifications through cryptographic checksums or filesystem-level audit hooks, is deployable today without specification changes, imposes low runtime overhead, and does not restrict any legitimate Skill behavior, since no legitimate Skill needs to modify the agent's core memory files without user awareness. Prevention via read-only mounting of configuration files during normal operation, with explicit user confirmation required for any modification, would prevent the CVE-2025-59536 class of attacks and is implementable within existing agent runtime architectures. The usability cost is a confirmation prompt for the small fraction of Skills that legitimately modify configuration.

\textbf{Against Multi-Agent Propagation (T7).} Prompt infection turns on downstream agents treating an inter-agent channel as they treat their own configuration. Trust boundary enforcement between agents, processing inter-agent messages at user level rather than at the level a Skill's own instructions occupy, would prevent propagation of adversarial instructions with elevated authority. This reform is architecturally feasible but requires coordination across agent frameworks, which currently lack standardized trust models for inter-agent communication~\cite{yu2025survey}. The utility cost is bounded, since the reform affects only Skills deployed in multi-agent configurations and does not restrict single-agent Skill functionality.

\subsection{Open Challenges}

The defense directions identified above reveal a set of open research challenges that must be addressed to establish Agent Skills security as a mature research area. We identify challenges that are not resolved by existing techniques.

\textbf{C1: Natural Language Instruction Analysis.} The fundamental security analysis challenge for Agent Skills is determining the behavioral envelope of a natural language instruction file. Unlike code, which can be statically analyzed for API calls, control flow, and data dependencies, a \texttt{SKILL.md} instructions body expresses behavior in natural language that is interpreted dynamically by a language model. There is currently no formal framework for specifying the intended behavior of a Skill, no static analysis tool that can reliably characterize what a Skill will direct the agent to do across the space of possible user requests, and no verification methodology for establishing that a Skill's actual behavior conforms to its declared description. Developing such a framework is a prerequisite for meaningful security auditing of Skill content.

\textbf{C2: Consent Model Design.} The consent gap identified in T2.1 is a consequence of the current all-or-nothing trust model, but replacing it with a fine-grained permission system introduces its own challenges. Permission prompts that are too fine-grained cause approval fatigue, leading users to approve all requests without reading them, the same failure mode that undermined Android's permission model in its early iterations. Permission prompts that are too coarse fail to meaningfully constrain the attack surface. Designing a consent model that is both security-effective and usable for non-expert users, calibrated to the specific action space of LLM-based agents, is an open problem that requires interdisciplinary research spanning security, human-computer interaction, and agent system design.

\textbf{C3: Runtime Behavioral Monitoring.} Runtime detection needs a reference model of legitimate behavior for each Skill, which C1 identifies as unavailable. Monitoring nonetheless faces constraints that static analysis does not. It observes action sequences rather than declarations, must decide within the latency budget of an agent turn, and pays for false positives in interrupted work rather than in rejected submissions. Anomaly detection over action sequences, behavioral fingerprinting of known-good executions, and intent classification of agent actions have been proposed, and none has been validated at the scale of a deployed marketplace.

\textbf{C4: Supply Chain Integrity at Scale.} Current cryptographic supply chain security frameworks, such as in-toto~\cite{torres2019toto} and sigstore, provide provenance verification for compiled artifacts but do not address the specific properties of natural language instruction files. A Skill whose \texttt{SKILL.md} is cryptographically signed is not thereby safe, since the signature verifies authorship but not behavioral intent. Extending supply chain integrity frameworks to cover the semantic properties of natural language Skill content, and not only the syntactic integrity of the file, requires new verification primitives that do not yet exist. The problem extends past the package boundary. A signature over a Skill's own files says nothing about the packages it resolves at execution time or the documents it fetches at runtime, both of which Section~\ref{sec:incidents} shows to be the operative attack surface. This challenge is compounded by the open contribution model of Agent Skills marketplaces, in which the volume of submissions makes manual review at scale impractical.

\textbf{C5: Trust Propagation in Multi-Agent Systems.} The prompt infection scenario (T7.1) is a specific instance of a broader open problem: how should trust be propagated across agent boundaries in multi-agent systems? When an orchestrator agent delegates a subtask to a subagent, what authority should the subagent inherit? When a subagent returns a result to the orchestrator, how should the orchestrator evaluate the trustworthiness of that result? These questions do not have answers in the current agent security literature, and the Agent Skills framework provides no guidance. Formalizing a trust propagation model for multi-agent systems that is both secure and operationally practical has implications beyond Agent Skills.

\textbf{C6: Automated Skill Vetting.} Marketplace-scale security review of Agent Skills requires automated vetting tools that can assess the security properties of a Skill without requiring manual inspection of every submission. Existing tools such as Snyk's agent-scan~\cite{snyk2026agentscan} and Cisco's skill-scanner~\cite{cisco2026skillscanner} provide heuristic detection of known patterns. A Skill constructed to be malicious passed every scanner applied to it while reaching more than 26,000 agents, because at scan time its files did what they claimed and the payload was served from elsewhere~\cite{air2026story}. Developing automated vetting pipelines that combine static analysis of bundled scripts, semantic analysis of natural language instructions, dynamic analysis of Skill behavior in sandboxed environments, and provenance verification of declared dependencies, while maintaining low false positive rates to avoid blocking legitimate Skills, is an open engineering and research challenge.

\textbf{C7: Specification-Level Security Properties.} The Agent Skills specification, as currently defined, makes no security guarantees and imposes no security requirements on conforming implementations. Extending the specification to include mandatory capability declarations, dependency pinning requirements, and content integrity mechanisms would enable the ecosystem to enforce baseline security properties across all compliant agent runtimes. Specification design for security is not straightforward. Requirements that are too prescriptive may limit the expressiveness of Skills, while requirements that are too permissive fail to meaningfully constrain the attack surface. Designing a security-aware specification that preserves the framework's flexibility requires engagement from the security research community, the agent platform developers, and the Skill author community.

\textbf{C8: Cross-Implementation Conformance.} The specification defines optional fields without defining what a conforming runtime must do with them, and Section~\ref{sec:architecture:package} documents runtimes that diverge in exactly this way. A user cannot presently determine, from a Skill and a runtime, which of that Skill's declarations will take effect, and neither can a tool auditing on the user's behalf. Establishing conformance requirements for security-relevant fields, together with a test suite against which runtimes could be checked, is a prerequisite for any security property that depends on a declaration being honored. The problem is organizational as much as technical, since the specification is now governed by a foundation whose members ship independent implementations, and conformance requirements bind only to the extent that those members adopt them.

\subsection{Recommendations}

The findings of this analysis carry practical implications for the four principal stakeholder groups in the Agent Skills ecosystem.

\textbf{For Skill authors.} Authors should treat \texttt{SKILL.md} files with the same security discipline applied to code. Bundled scripts should declare pinned dependencies with version-locked lockfiles, and any external URLs referenced in instructions should be validated against an explicit allowlist. Authors should avoid instructions that direct agents to read files or make network requests beyond the declared scope of the Skill, and should document the full set of filesystem and network operations the Skill may perform to enable users to make informed installation decisions.

\textbf{For marketplace operators.} Marketplace operators should implement mandatory security review pipelines that combine static analysis of bundled scripts, semantic analysis of \texttt{SKILL.md} instructions for injection patterns and overbroad capability claims, and provenance verification of declared dependencies. Namespace governance mechanisms, enforcing uniqueness constraints and binding namespaces to verified publisher identities, are necessary to prevent typosquatting at scale. Reputation systems should implement anomaly detection for coordinated inflation patterns and should weight verified installation outcomes over raw download metrics. Where a review has been performed, operators should be explicit about what it establishes, since a catalog that presents scanned Skills as vetted asserts more than a scan supports.

\textbf{For agent platform developers.} Platform developers should give the existing optional frontmatter fields defined enforcement semantics, so that a declaration binds every conforming runtime rather than being honored by some and discarded by others, and should extend the specification to require dependency pinning in bundled scripts and to implement version-bound trust grants that bind the trust relationship to a specific content hash of the Skill at installation time. Runtime sandboxing of bundled script execution, with explicit capability grants for filesystem access, network access, and subprocess invocation, should be a baseline requirement rather than an optional feature. Memory and configuration modifications should require explicit user confirmation and be logged to a tamper-evident audit trail.

\textbf{For enterprise users.} Enterprise users deploying Agent Skills in production environments should treat Skill installation as a software dependency management problem, applying the same governance processes used for open source package adoption: inventory tracking, security review before installation, ongoing monitoring for content changes, and incident response procedures for detected compromise. Purpose-built behavioral monitoring that observes agent network activity and filesystem access patterns, alerting on behavior inconsistent with the active Skill's declared purpose, should be considered a baseline operational requirement in environments where agents have access to sensitive codebases or credentials.

\section{Related Work}\label{sec:related}

The architectural properties of Agent Skills, analyzed in Section~\ref{sec:architecture}, give rise to a threat landscape that existing security research has addressed only partially. We survey the related work and identify the gap each leaves with respect to Agent Skills security.

\textbf{LLM Adversarial Robustness.} A large body of work studies the vulnerability of language models to adversarial inputs, including jailbreaks that elicit policy-violating outputs~\cite{wei2023jailbroken,zou2023universal,chao2025jailbreaking,mehrotra2024tree}, prompt injection attacks that hijack model behavior through adversarial instructions~\cite{perez2022ignore,jia2025promptlocate}, and model extraction and membership inference attacks targeting the underlying weights~\cite{carlini2021extracting}. This line of work treats the model as the primary attack surface and focuses on perturbations to model inputs or training data. The threats we identify arise from the extension mechanism rather than from the model itself, and the attacker model encompasses actors who control Skill content rather than actors who craft adversarial inputs. These techniques address neither trust establishment, nor permission scoping, nor supply chain compromise.

\textbf{Extension Ecosystem Security.} Browser extensions and IDE plugins resemble Agent Skills in three respects. They are contributed by third parties rather than by the platform vendor, they are loaded into a running application rather than compiled into it, and they obtain broad access to the user's environment once installed. Research on Chrome extensions has identified large-scale malicious and overprivileged extension populations~\cite{jagpal2015trends,kapravelos2014hulk}, and analogous work on VS Code extensions has found widespread suspicious behavior patterns~\cite{edirimannage2024developers}. Plugin review processes have been studied in the browser extension context, revealing systematic limitations of human-review-based vetting at scale~\cite{jagpal2015trends}. A closer precedent is the instruction file mechanism that AI coding assistants read from the project directory. Pillar Security showed that instructions concealed in such files using invisible Unicode characters were legible to the model but not to a human reviewer, and could direct GitHub Copilot and Cursor to emit backdoored code that the assistants did not mention in their responses~\cite{pillar2025rulesfile}. Agent Skills inherits these problems and adds the dimension of natural language instruction delivery. Unlike code-based extensions that can be statically analyzed for API calls and permission usage, a Skill's behavioral specification is written in natural language and interpreted dynamically by a language model, so that static analysis cannot characterize the full behavioral envelope.

\textbf{Software Supply Chain Security.} Software supply chain attacks have become a major threat vector, with package repository compromises delivering malware to large downstream populations~\cite{ladisa2023sok,ohm2020backstabber}. The SolarWinds~\cite{solarwinds}, XZ Utils~\cite{xz_backdoor}, and analogous incidents have driven extensive research on malicious package detection~\cite{ohm2020backstabber}, dependency confusion~\cite{birsan2021dependency}, and cryptographic provenance verification frameworks~\cite{torres2019toto}. Agent Skills marketplaces exhibit the core properties that make package repositories vulnerable, namely open submission, automated distribution, and large consumer populations, with the additional complication that malicious payload delivery requires neither a compromised binary nor injected executable code, since redirecting agent behavior requires only that a sentence be added to a text file. Unlike binary supply chain attacks, which can in principle be detected through cryptographic integrity checks on compiled artifacts, the behavioral effects of a malicious \texttt{SKILL.md} are indistinguishable from benign instruction drift without a ground-truth specification of intended agent behavior. The real-world ClawHavoc campaign~\cite{antiy2026clawhavoc,liu2026malicious} confirms that this attack surface is actively exploited at scale.

\textbf{LLM-based Agent Security.} A growing body of work addresses security threats specific to LLM-based agents. At the survey level, Gan et al.\ provide a comprehensive taxonomy of security, privacy, and ethics threats organized around threat sources and impacts~\cite{gan2024navigating}; He et al.\ catalog security and privacy issues with case studies across representative deployments~\cite{he2025emerged}; Yu et al.\ survey threats and countermeasures with particular attention to multi-agent trust dynamics~\cite{yu2025survey}; and Datta et al.\ offer a broad treatment covering evaluation methodologies and open challenges~\cite{datta2025agentic}. At the attack level, prompt injection represents the most studied threat vector. Greshake et al.~\cite{greshake2023not} introduced indirect prompt injection, demonstrating that adversarial instructions embedded in external content retrieved by an agent can hijack its behavior; subsequent work has formalized attack taxonomies, developed benchmarks~\cite{zhan2024injecagent,liu2024formalizing,debenedetti2024agentdojo}, and characterized the difficulty of robust mitigation under adaptive attackers~\cite{liu2024formalizing}. Defenses have explored both training-based and inference-time approaches, including structured query formats~\cite{chen2025struq} and privilege-based instruction hierarchies~\cite{wallace2024instruction}. Beyond prompt injection, Chen et al.\ introduced AgentPoison, targeting RAG-based agents by poisoning their long-term memory~\cite{chen2024agentpoison}, and Yang et al.\ formalized agent backdoor attacks embeddable through data poisoning or weight manipulation~\cite{yang2024watch}. The MCP security analysis of Hou et al.~\cite{hou2025mcp} identifies nine threat categories across the MCP lifecycle. Agent Skills nonetheless introduces architectural properties that have no MCP counterpart: the natural language instruction carrier, the absence of a typed interface contract, and the single-approval persistent trust model. Existing prompt injection defenses rely on a privilege boundary between trusted and untrusted content~\cite{wallace2024instruction,chen2025struq}, a boundary that Section~\ref{sec:discussion} shows Agent Skills does not provide.

\textbf{Agent Skills Security.} Work specific to Agent Skills falls into four groups. \emph{Attack demonstrations} established the tractability of Skill-borne injection. Schmotz et al.\ showed that a few lines of Markdown suffice to redirect agent behavior~\cite{schmotz2025trivial} and subsequently measured agent vulnerability to Skill file attacks across models~\cite{schmotz2026skillinject}; Jia et al.\ automated the generation of poisoned Skills through instruction rewriting and helper scripts~\cite{jia2026skillject}. \emph{Measurement studies} quantified prevalence: Liu et al.\ scanned 42,447 Skills and reported that 26.1\% carry at least one vulnerability~\cite{liu2026agent}, and their companion study characterized malicious Skills observed in deployment~\cite{liu2026malicious}. Holzbauer et al.\ analyzed the ecosystem with attention to the repository context in which Skills are published~\cite{holzbauer2026context}. \emph{Defenses and benchmarks} have begun to appear, including formal treatments of the skill dependency graph~\cite{bhardwaj2026formal} and evaluation suites for detecting malicious Skills~\cite{wang2026elementary}. \emph{Systematizations} closest to the present work include Xu and Yan, who survey Skill architecture and acquisition alongside security~\cite{xu2026agentskills}, and OWASP's catalogue for agentic Skills, which enumerates risks for practitioner use~\cite{owasp2026skills}. Table~\ref{tab:novelty} compares our coverage with theirs and with the MCP analysis of Hou et al.

\begin{table}[t]
\centering
\small
\setlength{\tabcolsep}{4pt}
\renewcommand{\arraystretch}{1.3}
\caption{Coverage of our threat categories by the closest prior and concurrent systematizations. \newmark~treated as a distinct category with an identified cause; \partialmark~mentioned or covered in part; \nomark~not addressed.}
\label{tab:novelty}
\begin{tabular}{
  >{\raggedright\arraybackslash}m{\dimexpr 0.4\linewidth-4.0\tabcolsep\relax}
  >{\centering\arraybackslash}m{\dimexpr 0.16\linewidth-1.6\tabcolsep\relax}
  >{\centering\arraybackslash}m{\dimexpr 0.16\linewidth-1.6\tabcolsep\relax}
  >{\centering\arraybackslash}m{\dimexpr 0.16\linewidth-1.6\tabcolsep\relax}
  >{\centering\arraybackslash}m{\dimexpr 0.12\linewidth-1.2\tabcolsep\relax}
}
\toprule
\textbf{Dimension}
  & \textbf{\shortstack{Hou et al.\\(MCP)~\cite{hou2025mcp}}}
  & \textbf{\shortstack{Xu and\\Yan~\cite{xu2026agentskills}}}
  & \textbf{\shortstack{OWASP\\Skills~\cite{owasp2026skills}}}
  & \textbf{Ours} \\
\midrule

\rowcolor{groupbg}
\multicolumn{5}{l}{\textit{\small Threat Categories}} \\

T1 Supply chain compromise & \newmark & \newmark & \newmark & \newmark \\
\rowcolor{rowgray}
T2 Consent abuse & \partialmark & \partialmark & \partialmark & \newmark \\
T3 Prompt injection & \newmark & \newmark & \newmark & \newmark \\
\rowcolor{rowgray}
T4 Code execution & \partialmark & \newmark & \newmark & \newmark \\
T5 Data exfiltration & \newmark & \newmark & \newmark & \newmark \\
\rowcolor{rowgray}
T6 Persistence & \nomark & \partialmark & \newmark & \newmark \\
T7 Multi-agent propagation & \nomark & \nomark & \partialmark & \newmark \\

\midrule
\rowcolor{groupbg}
\multicolumn{5}{l}{\textit{\small Organizing Dimensions}} \\

Lifecycle decomposition & \newmark & \partialmark & \nomark & \newmark \\
\rowcolor{rowgray}
Specification versus implementation origin & \nomark & \nomark & \nomark & \newmark \\
Evidence per scenario & \nomark & \nomark & \partialmark & \newmark \\
\rowcolor{rowgray}
Mitigation feasibility per scenario & \nomark & \nomark & \partialmark & \newmark \\

\bottomrule
\end{tabular}
\end{table}

\section{Conclusion}
\label{sec:conclusion}
This paper presents a systematization of the security of Agent Skills. We define a threat model for the framework, decompose its attack surface across a four-phase lifecycle, construct a threat taxonomy comprising seven categories and nineteen scenarios grounded in real-world evidence, and map seven documented incidents onto it. The most severe threats in the Agent Skills ecosystem are not addressable through incremental mitigations: prompt injection resists detection without a formal behavioral specification that does not exist, the consent gap requires architectural reform to the trust model, and supply chain integrity cannot be ensured by signing a Skill's own files, because the code it executes and the instructions it follows may both be resolved afterwards. As natural language capability specification becomes the dominant paradigm in agentic AI, we hope this work provides a foundation for the security research and engineering effort that safe broad deployment will require.
\bibliographystyle{ACM-Reference-Format}
\bibliography{software}

\appendix

\section{Assignment Principles and Per-Cell Evidence}\label{app:rubric}

This appendix records the basis for each assignment in Table~\ref{tab:comparison}. Four principles govern the assignments.

First, each cell is assessed against the mechanism's own published specification and vendor documentation, not against an idealized or a worst-case deployment.

Second, an amplifying design property is counted once, at the dimension that produces it. The specification's positioning of Skill instructions at operator level amplifies several exposures, and we record it under Trust Model rather than repeating it across every dimension it touches.

Third, a control that the design enforces is marked \cmark, and one whose enforcement depends on correct client behavior is marked \warn. The absence of a control is marked \xmark. The absence of a control combined with a design property that makes the exposure default, invisible, or elevated in authority is marked \crit.

Fourth, a dimension is classified as implementation-determined when the standard leaves it unconstrained. ChatGPT Plugins receives no range on these dimensions because its manifest model fixes execution on provider infrastructure, which removes the latitude that MCP and Agent Skills leave to individual runtimes.

\subsection*{Specification-Determined Properties}

\small
\begin{longtable}{>{\raggedright\arraybackslash}p{\dimexpr .21\linewidth-1.26\tabcolsep\relax} >{\centering\arraybackslash}p{\dimexpr .08\linewidth-.48\tabcolsep\relax} p{\dimexpr .71\linewidth-4.26\tabcolsep\relax}}
\caption{Per-cell evidence for specification-determined properties.}
\label{tab:rubric-spec} \\

\toprule
\textbf{Mechanism} & \textbf{Level} & \textbf{Basis} \\
\midrule
\endfirsthead

\multicolumn{3}{c}{\tablename~\thetable{} (continued)} \\
\toprule
\textbf{Mechanism} & \textbf{Level} & \textbf{Basis} \\
\midrule
\endhead

\multicolumn{3}{l}{\textit{Instruction Carrier}} \\*
\midrule
ChatGPT Plugins & \cmark & The capability is conveyed as an OpenAPI manifest. The document is machine-parseable and enumerates endpoints with typed parameters, so an author cannot express behavioral directives outside the schema~\cite{openai2023plugins}. \\
MCP & \warn & Tools, resources, and prompts are declared with structured schemas, but the human-readable description fields accompanying each tool are model-facing prose and are read by the model when selecting operations~\cite{hou2025mcp}. \\
Agent Skills & \xmark & The instructions body of \texttt{SKILL.md} is unstructured Markdown interpreted directly by the model. The specification imposes no schema on its content~\cite{agentskills2025spec,anthropic2025skills}. \\
\midrule

\multicolumn{3}{l}{\textit{Data/Instruction Boundary}} \\*
\midrule
ChatGPT Plugins & \cmark & The manifest and the API response are distinct artifacts. Response payloads return to the model within a typed envelope and cannot alter the set of operations the model may invoke~\cite{openai2023plugins}. \\
MCP & \warn & Typed invocation separates the operation from its parameters, but tool results and resource contents enter the context as text alongside the model's instructions~\cite{hou2025mcp}. \\
Agent Skills & \crit & Author-supplied instructions and content retrieved at runtime occupy the same natural language medium, and the specification defines no marker distinguishing them. Retrieved content enters the same context region that already holds the Skill's instructions~\cite{anthropic2025skills,schmotz2025trivial}. \\
\midrule

\multicolumn{3}{l}{\textit{Capability Declaration}} \\*
\midrule
ChatGPT Plugins & \cmark & The manifest enumerates every operation the plugin exposes, and the enumeration binds. Operations absent from the manifest are unreachable~\cite{openai2023plugins}. \\
MCP & \warn & Servers declare their tools with typed schemas, which bounds what a client may invoke. The declaration does not cover the filesystem or network resources the server itself accesses during execution~\cite{hou2025mcp}. \\
Agent Skills & \xmark & The specification requires only \texttt{name} and \texttt{description}. The \texttt{allowed-tools} field is optional and carries no defined enforcement semantics, and conforming runtimes differ in whether they act on it~\cite{agentskills2025spec,anthropic2026skillsbestpractices}. No field declares the files a Skill reads or the endpoints it contacts. \\
\midrule

\pagebreak
\multicolumn{3}{l}{\textit{Trust Model}} \\
\midrule
ChatGPT Plugins & \cmark & Activation is per conversation. The user selects which plugins are enabled for a given session, and the granted authority is bounded by the manifest~\cite{openai2023plugins}. \\
MCP & \xmark & Approval is granted at server installation, and the protocol mandates no per-action confirmation. Whether a client prompts before each tool invocation is left to the client~\cite{hou2025mcp}. \\
Agent Skills & \crit & A single installation approval grants authority that persists across sessions, covers all actions the Skill may subsequently direct, and is not rebound when the Skill's content changes. The specification additionally positions an activated Skill's instructions at operator level~\cite{anthropic2025skills,cherny2025medusalocker}. \\
\bottomrule
\end{longtable}

\subsection*{Implementation-Determined Properties}

\begin{longtable}{>{\raggedright\arraybackslash}p{\dimexpr .21\linewidth-1.26\tabcolsep\relax} >{\centering\arraybackslash}p{\dimexpr .08\linewidth-.48\tabcolsep\relax} p{\dimexpr .71\linewidth-4.26\tabcolsep\relax}}
\caption{Per-cell evidence for implementation-determined properties.}
\label{tab:rubric-impl} \\

\toprule
\textbf{Mechanism} & \textbf{Level} & \textbf{Basis} \\
\midrule
\endfirsthead

\multicolumn{3}{c}{\tablename~\thetable{} (continued)} \\
\toprule
\textbf{Mechanism} & \textbf{Level} & \textbf{Basis} \\
\midrule
\endhead

\multicolumn{3}{l}{\textit{Execution Locus}} \\*
\midrule
ChatGPT Plugins & \cmark & Execution occurs on the provider's infrastructure. No plugin code runs on the user's machine~\cite{openai2023plugins}. \\
MCP & \rng{\warn}{\xmark} & Servers may be reached over HTTP, in which case execution is remote, or launched over stdio, in which case the server runs as a local user process~\cite{hou2025mcp}. \\
Agent Skills & \rng{\warn}{\xmark} & Skills execute within a managed environment on the Claude API and on the user's host under Claude Code~\cite{pluto2026extensions}. \\
\midrule

\multicolumn{3}{l}{\textit{Runtime Isolation}} \\*
\midrule
ChatGPT Plugins & \cmark & Isolation follows from the execution locus. The user's machine is not a target of plugin code~\cite{openai2023plugins}. \\
MCP & \rng{\warn}{\xmark} & Remote servers are isolated from the user's host by construction. Local servers run without isolation unless the client supplies it, and the protocol requires none~\cite{hou2025mcp}. \\
Agent Skills & \rng{\warn}{\crit} & Claude Code offers filesystem and network isolation modes, but they are opt-in and were absent when the format was introduced~\cite{anthropic2026sandbox}. Where isolation is absent, the exposure is amplified by the loading model. A bundled script's source never enters the context window, so the agent acts on the script's output without visibility into its behavior~\cite{anthropic2025skills}. \\
\midrule

\multicolumn{3}{l}{\textit{Permission Scope}} \\*
\midrule
ChatGPT Plugins & \cmark & The reachable resource set is the set of endpoints named in the manifest~\cite{openai2023plugins}. \\
MCP & \rng{\warn}{\xmark} & Typed schemas bound what a client may request of a server. A locally launched server inherits the user's privileges and its internal resource access is unconstrained by the protocol~\cite{hou2025mcp}. \\
Agent Skills & \rng{\warn}{\crit} & Runtimes that implement permission prompts or tool allowlists constrain the reachable set. Where they do not, a Skill operates at the user's privilege level with no declared scope against which any access could be judged out of bounds~\cite{anthropic2025skills,pluto2026extensions}. \\
\bottomrule
\end{longtable}

\subsection*{Distribution and Ecosystem Governance}

\begin{longtable}{>{\raggedright\arraybackslash}p{\dimexpr .21\linewidth-1.26\tabcolsep\relax} >{\centering\arraybackslash}p{\dimexpr .08\linewidth-.48\tabcolsep\relax} p{\dimexpr .71\linewidth-4.26\tabcolsep\relax}}
\caption{Per-cell evidence for distribution and ecosystem governance.}
\label{tab:rubric-eco} \\

\toprule
\textbf{Mechanism} & \textbf{Level} & \textbf{Basis} \\
\midrule
\endfirsthead

\multicolumn{3}{c}{\tablename~\thetable{} (continued)} \\
\toprule
\textbf{Mechanism} & \textbf{Level} & \textbf{Basis} \\
\midrule
\endhead

\multicolumn{3}{l}{\textit{Marketplace Review}} \\*
\midrule
ChatGPT Plugins & \cmark & Publication in the plugin store required passing a review process combining automated and human checks~\cite{customgpt2023review}. \\
MCP & \xmark & Distribution is decentralized. Servers are installed directly from source repositories, and no review gate stands between publication and installation~\cite{hou2025mcp}. \\
Agent Skills & \xmark & The major community marketplaces impose no mandatory review before publication~\cite{snyk2026toxicskills}. Curated registries that scan and sign Skills have since appeared~\cite{nvidia2026verified}, but they are additional channels rather than a requirement of the ecosystem. \\
\bottomrule
\end{longtable}
\normalsize

\noindent
Authorship Complexity is reported on a separate scale and is not a security level. ChatGPT Plugins is rated High because authorship required writing an OpenAPI specification and operating a remote endpoint. MCP is rated Medium because authorship required implementing a protocol-compliant server. Agent Skills is rated Low because a single Markdown file with two metadata fields constitutes a publishable Skill~\cite{agentskills2025spec}.

\end{document}